\documentclass[preprinted,showpacs,preprintnumbers,amsmath,amssymb,superscriptaddress,floatfix]{revtex4}
\usepackage{graphicx}
\usepackage{color}
\usepackage{dcolumn}
\usepackage{bm}
\usepackage[latin1]{inputenc}

\begin{document}
\preprint{Preprint -- H. S. Ruiz and A. Bad\'{\i}a-Maj\'{o}s}
\title{Strength of the phonon-coupling mode in $La_{2-x}Sr_{x}CuO_{4}$, $Bi_{2}Sr_{2}CaCu_{2}O_{8+x}$ and $YBa_{2}Cu_{3}O_{6+x}$: An estimation from the ARPES-nodal measurements.}
\author{H. S. Ruiz}
\email[Electronic address: ]{hsruizr@unizar.es}
\affiliation{Departamento de F\'{\i}sica de la Materia
Condensada--I.C.M.A., Universidad de Zaragoza--C.S.I.C., Mar\'{\i}a
de Luna 1, E-50018 Zaragoza, Spain}

\author{A. Bad\'{\i}a\,--\,Maj\'os}
\affiliation{Departamento de F\'{\i}sica de la Materia
Condensada--I.C.M.A., Universidad de Zaragoza--C.S.I.C., Mar\'{\i}a
de Luna 1, E-50018 Zaragoza, Spain}

\begin{abstract}
Despite the intensive efforts for determining the mechanism that causes high-temperature superconductivity in copper oxide materials, no consensus on the pairing mechanism has been reached. Recent advances in high resolution angle-resolved photoemission spectroscopies have suggested that a sizeable electron-phonon coupling exists as the principal cause for kinks in the dispersion relations (energy versus wave vector) of the electronic states. Here, we report on a systematic study about the influence of the electron-phonon coupling parameter ``$\lambda$'' in the electronic quasiparticle dispersions along the nodal direction for $La_{2-x}Sr_{x}CuO_{4}$, $Bi_{2}Sr_{2}CaCu_{2}O_{8+x}$ and $YBa_{2}Cu_{3}O_{6+x}$. Information about the dressing of the charge carriers, i.e., on the enhancement of the effective mass and the strength of the coupling mode, is obtained as a function of the doping concentration, temperature, momentum and energy from the kink dispersion in the (0-0)-$(\pi,\pi)$ direction of momentum-space avoiding the complications of the \textit{d-}wave superconducting gap. Our analysis shows a remarkable agreement between theory and experiment for different samples and at different doping levels. This includes our recently introduced theoretical model to adjust the experimental data of the fermionic band dispersion, emphasising the necessary distinction between the general electron mass-enhancement parameter $\lambda^{*}$ and the conventional electron-phonon coupling parameter $\lambda$. In LSCO, the coupling constant $\lambda$, calculated consistently with the nodal kink dispersions, reproduces the observed critical temperatures $T_{c}$, the gap ratio $2\Delta_{0}/k_{B}T_{c}$, and other parameters which have been studied from several equations. It will be concluded that the strong renormalisation of the band structure can be explained in terms of the phonon coupling mode, and must therefore be included in any microscopic theory of superconductivity, even for those materials in which the contribution to the pair formation can be less dominant. Nevertheless, it seems unavoidable to consider additional mechanisms that justify the higher critical temperatures observed in BSCCO and YBCO samples.
\end{abstract}


\maketitle

\section{Introduction}
\label{Sec:Intro}


One of the most relevant milestones in solid state physics has been undoubtedly the discovery of the high superconducting transition temperature ($T_{c}$) in LaBaCuO ceramics, prompting an intense activity in the field of synthesis and characterisation of copper oxide materials (cuprates)~\cite{Bednorz86}. The chemical composition of cuprates is well characterised by a layered crystal structure with one or more CuO$_{2}$ planes per unit cell, which are responsible for the low lying electronic structure, and also presumably, the key for understanding the quasiparticle properties involved in the mechanism of superconductivity, i.e.: the origin of the coupling mode which binds two electrons (holes) in the formation of Cooper pairs. Nevertheless, despite over two decades of experimental and theoretical efforts to elucidate the pairing mechanism and to understand the many-body properties influencing the electron dynamics in cuprate superconductors, there is no consensus on the strength of the electron correlations and the nature of the boson coupling modes. Nowadays, with the advent of improvement in instrumental resolution, sample quality, as well as theoretical development, the relevance of electron-boson coupling mechanisms on the electronic dynamics of cuprates is more and more accepted, because the many body effects are supposed to be particularly strong~\cite{Hufner07}. In this paper, we will review recent progress in the understanding of electrons interacting with bosonic modes, such as phonons, and discuss the results of recent high-resolution angle-resolved photoemission spectra (ARPES) on cuprate superconductors with the aim of studying its possible connection with the occurrence of high-$T_{c}$. In order to  understand the relevance of the ARPES technique, we recall that the photoemission process results in both an excited photoelectron and a photohole in the final state. Related to this observation, it becomes a useful probe of the related scattering mechanisms contributing to the electrical transport in different materials. Unlike other probes of the transport properties, the ARPES technique has the advantage of momentum resolving. Along this line, we want to note that the single-particle scattering rate measured in ARPES is not identical to the scattering rate measured in transport studies themselves. Nonetheless, direct proportionality between them has been established~\cite{Kulic00,Smith01}. 

In conventional metals, the phonon coupling mode has long been recognised as the main mechanism involved in the superconducting properties, and the strength of this interaction essentially determines the value of $T_{c}$. On the other hand, in the high-temperature superconductors (HTSC) the experimentally determined $d-$wave pairing introduces considerable complications in the theory even when other coupling modes are considered~\cite{Kulic04}. Fortunately, with the appearance of a new era of analysers with improved resolution both in energy and momentum, in angle-resolved photoemission spectroscopy (ARPES) as well as in inelastic neutron scattering (INS) and x-ray scattering (IXS) experiments, the controversy on the influence of the anisotropic character of the superconducting gap in the electron properties can be avoided, by analysing preferential directions within the $CuO_{2}$ planes or nodal directions. This provides a smart solution if one is merely interested in identifying the energy modes~\cite{Devereaux04}. In fact, one of the most telling manifestations of the electron-phonon coupling is a mass renormalisation of the electronic dispersion at the energy scale associated with the phonons. This renormalisation effect is directly observable in the ARPES measurements as a low-energy excitation band in the dispersion curves of photoemitted electrons, known as \textit{kink}~\cite{Lanzara01,Zhou03}. This feature, so far universal in HTSC, has been regarded as a signature of the strength of the boson mechanism which causes the pair formation in the superconducting state. In fact, all the interactions of the electrons which are responsible for the unusual normal and superconducting properties of cuprates are believed to be represented in this anomaly~\cite{Zhou07}. This has prompted an intense debate about the nature of the coupling mode involved in the density of low-energy electronic excited states in the momentum-energy space of cuprates, and its influence on the emergence of the superconducting state~\cite{Devereaux04,Lanzara01,Zhou03,Zhou07,Zhou02,Zhou05,Kordyuk06,Xiao07,Takahashi07,Gweon04,Douglas07,Johnson01,Borisenko06,Zhang08,Graf08,Reznik08,Giustino08,Park08,Chang08}.  

So far, the origin of high-temperature superconductivity and the nature of the bosons involved remains controversial mainly because experiments can only be used to determine an approximate energy of the mode and this energy is close to both the optical phonons \cite{Lanzara01,Zhou03,Zhou05,Zhang08,Graf08,Reznik08} and magnetic excitations \cite{Kordyuk06,Xiao07,Giustino08,Dahm09,Graf07,Terashima06}. Related to this, the energy scales of the optical phonons in the $\texttt{CuO}_{2}$ planes are similar for electron and hole doped cuprates, while the magnetic mode for electron doped HTSC is found to be much smaller~\cite{Wilson06,Zhao07}. Along this line, recently S. R. Park \textit{et al.}~\cite{Park08} have demonstrated that the magnetic resonance mode can not explain the ARPES spectra in several electron doped HTSC systems showing a clear support for the electron-phonon coupling. In addition, J. Graf \textit{et al.}~\cite{Graf08} have recently reported the first evidence of an anomalous dispersion of the Cu-O bond stretching phonon mode in a Bi cuprate supporting the idea that strong electron energy dispersion measured by ARPES corresponds to the Cu-O bond stretching phonon mode. On the other hand, considering that the magnetic resonance has not been detected in the single layer Bi2201~\cite{Sato03}, magnetic modes should be ruled out as a general mechanism within a controllable theory of strong correlations. 

In an effort to clarify the influence of the phonon coupling mode (either weak or strong), in this work we are going to analyse the influence of the so-called electron-phonon coupling parameter on the electronic dispersion relations in several cuprate compounds.


The paper is organised as follows. In Sec.~\ref{Sec_1}, we put forward some details about the conventional electron-phonon coupling theory focusing on the determination of the complex electron self-energy, and evaluating the influence of the electron-phonon spectral density calculated by different methods available in the literature on the superconducting properties of some cuprates (e.g., LSCO, YBCO, BSCCO). We concentrate on the use of approximate formulas to determine the superconducting transition temperature $T_{c}$, the ratio gap $2\Delta_{0}/k_{B}T_{c}$, and the zero temperature gap $\Delta_{0}$. The physical interpretation of the underlying approximations is also focused on. In Sec.~\ref{Sec_2} we present a systematic study on the influence of the electron-phonon coupling parameter in the electron quasiparticle dispersion relations unravelling possible new physics. Comparison with the electronic dispersion curves measured by ARPES at the nodal point will be emphasised for the three families of cuprates. An empirical equation is incorporated for determining $\lambda$ from the doping level in LSCO which may be of interest for other works. Finally, Sec.\ref{Sec:Conclusions} is devoted to
discussing our results. The relevance of the electron-phonon coupling
mechanism for the interpretation of the electron dynamics in HTSC will be concluded. Comparison with recent published material will be emphasised.

\section{Elements of electron-phonon coupling theory}
\label{Sec_1}


The purpose of this section is a brief reexamination of the conventional theory for electron-phonon coupling in condensed matter, with the aim of obtaining an appropriate interpretation of the so-called spectral function. This will envisage to extract some general consequences and then, check them from the direct interpolation between theory and experiment.


Recall that, nowadays the simplest interpretation of microscopic superconductivity remains to be the Bardeen, Cooper, and Schrieffer's \cite{Bardeen57} theory (BCS) which is based on the condensation of pairs of electrons (holes) into a spin singlet induced by an isotropic electron-phonon coupling. The formation of pairs gives rise to an excitation energy gap $\Delta$ in the electron density of states across the Fermi level ($E_{F}$) and its magnitude reflects the strength of the pairing interaction. In BCS the pairing function $\Delta$ has an isotropic $s$-\textit{wave} symmetry, and the actual reduced gap value $2\Delta_{0}/k_{B}T_{c}$ is used to classify a system as a weak- or strong coupling superconductor by comparison to the mean-field value 3.51. A significant accumulation of microscopic information on the fine structure found in tunnelling experiments at voltages above the forbidden energy gap was the most direct proof of the electron-phonon interaction as the coupling mechanism for superconductivity in conventional metals. However, along with the confirmation of the BCS predictions, the experimental results also indicated deviations from the weak-coupling BCS theory through the detection of large gaps as compared to the mean-field value. For the new generation of HTSC the detailed structure in the measured quasiparticle density of states, also revealed the need of one controllable strong-coupling theory. Moreover, many of the new families of superconductors discovered after the appearance of the cuprate compounds also exhibit highly anisotropic gaps, and transport and magnetic properties that cannot be directly explained by the simple BCS theory \cite{Schrieffer07}. At present, there is no consensus on the microscopical theory involved in the HTSC materials and the influence of the pairing mechanism on the superconducting gap and the higher $T_{c}$. Nevertheless, assuming that the quasiparticle dynamics is governed by some electron-boson coupling which predominates both in the normal state properties and in the superconducting state, a wide number of properties can be satisfactorily explained \cite{Kulic04,Carbotte90}. Thus, having made these general considerations on the many-body problem underlying in the electron-phonon coupling theory, we are going to expose some technical aspects to be considered in the analysis of ARPES data and their relation with the superconducting properties.

\subsection{The Migdal-Eliashberg approach: spectral functions}


In the diagrammatic language, the above mentioned physical properties are formulated within the framework of the Fermi-liquid theory, where electron-like quasiparticles populate bands in energy-momentum space up to the cut-off at the Fermi energy. Lattice vibrations couple to electrons because the displacements of atoms from their equilibrium positions alter the band dispersions, either lowering or raising the total electron quasiparticle energy. 


Electron correlations are responsible of the formation of quasiparticles which are well defined near the Fermi level. There, the so called vertex corrections can be a \textit{priori} neglected because these can be shown to be reduced by the ratio between the phonon frequency ($0-100~meV$) and  $E_{F}$ ($\sim1-10~eV$). In other words, the presence of strong electron correlations mediated by the electron-phonon interaction is avoided and the multi-phonon excitations are reduced to the single-loop approximation or Migdal-Eliashberg (ME) approach (Fig.~\ref{Fig_1}~d). In ordinary metals, this simple picture for the phonon-mediated interaction between electrons, or electron-phonon coupling, has long been known to be the pairing mechanism responsible for superconductivity. On the other hand, in the HTSC it has been suggested that other collective excitation modes mediate the pairing since the superconducting transition temperatures $T_{c}$ are much higher than those of conventional superconductors. Nonetheless we want to emphasise that the feasibility of the ME approach in the HTSC is an issue still open. In fact, the recent observation of similar renormalisation effects in the energy-momentum space of HTSC has raised the hope that the mechanism of high-$T_{c}$ superconductivity may finally be resolved, thus reviving the interest on the diagrammatic approach \cite{Hufner07,Lanzara01,Zhou03,Zhou07,Park08,Chang08,Ruiz09}.


Let us go into some more detail. The thermodynamic properties of the solid can be evaluated from the thermal Green's-function formalism where it is assumed that the electron-phonon interaction can be introduced by the relation
%
\begin{equation}
G^{-1}(k,i\omega_{n})=G_{0}^{-1}(k,i\omega_{n})-\Sigma(k,i\omega_{n})
\label{eq_1}
\end{equation}
with $G_{0}^{-1}$ related to the bare electron energy, i.e., the non-interacting electron Green's function ($G_{0}(k,i\omega_{n})=(i\omega_{n}-\varepsilon_{k})^{-1}$) and
$i\omega_{n}$ standing for the so-called ``imaginary Matsubara
frequencies'' \cite{Allen78}.  


Technically, the lowest order Feynman graph approximation for the electron phonon interaction can be defined on a set of orthonormal functions (harmonic Fermi surfaces) that allow to consider the complexities of the $d-$band electron structure and phonons from the ``angular'' and ``energy'' components of the phase space~\cite{Allen76}, i.e.: $k$-space is represented in terms of its harmonic components $(J,\varepsilon)$. Thus, in terms of this set
%
%
\begin{eqnarray}
\Sigma_{J}(k,i\omega_{n})=&&T\sum_{J',\nu}\int_{0}^{\infty}d\varepsilon'\frac{N(\varepsilon')}{N(0)}\int_{0}^{\infty}d\nu\alpha^{2}\emph{F}(J,J',\varepsilon,\varepsilon',\nu)\left(\frac{2\nu}{\omega_{\nu}^{2}+\nu^{2}}\right)\nonumber\\&&\times G_{J'}(\varepsilon',i\omega_{n}-i\omega_{\nu})
\label{eq_2}
\end{eqnarray}
%
where $N(0)=\sum_{k}\delta(\varepsilon_{k})$ is the electronic density states on the Fermi surface, and the electron-phonon spectral density $\alpha^{2}\emph{F}(J,J',\varepsilon,\varepsilon',\nu)$ represents a measure of the effectiveness of the phonons of frequency $\nu$ in the scattering electrons from $k(J,\varepsilon)$  to $k'(J',\varepsilon')$. Up to this point, only the harmonic approximation to the phonon propagator $D(k-k',\nu)=2\nu/(\omega_{\nu}^{2}+\nu^{2})$ has been considered. Nevertheless, we recall that the actual form of Eq.~(\ref{eq_2}) is cumbersome, and requires of an accurate determination of the electron-phonon spectral density from theoretical ab-initio calculations of the band structure of the crystal. Then, the ME approach is invoked. In order, to simplify Eq.~(\ref{eq_2}) one assumes that it is possible to neglect the dependence on the energy surfaces ($\varepsilon,\varepsilon'$) of the $N(\varepsilon')\alpha^{2}F(J,J',\varepsilon,\varepsilon',\nu)$ function. This allows us to omit the processes, violating the Born-Oppenheimer adiabatic theorem contained within the high order graphs. This is the ME ``approach''. Motivations for the use of this approach have been broadly discussed in the past (see Refs.~\cite{Carbotte90} \& \cite{Allen82}).


So far, the self-energy has been described in the orthonormal space of ``Fermi surface harmonics'' which is to be transformed in the phase-space under consideration for \textbf{k}-space rotations according to their irreducible representations of the point group of crystal. Formally, this means that the spectral function $\alpha^{2}F(JJ',\nu)$ is to be diagonal. It must be emphasised that this diagonal representation for the normal state will continue to hold in the superconducting state for the isotropic Cooper pairing or $s$-\textit{wave} gap. However, other pairing schemes which break rotation symmetry are, in principle, possible. Thus, taking advantage that the ARPES measurements at the nodal direction, are not influenced by the anisotropy of the superconducting gap, we will refer to a (nondirectional) isotropic quasiparticle spectral density, defined as the double average over the Fermi surface of the electron-phonon spectral density $\alpha^{2}F(\textbf{k},\textbf{k}',\nu)$; i.e., 
%
%
\begin{eqnarray}
    \alpha^{2}F(\nu)=\frac{1}{N(0)}\sum_{\textbf{kk'},j} \mid
    g_{\textbf{kk'}}^{j}\mid^{2}\delta(\nu-\nu_{\textbf{k}-\textbf{k'}}^{j})
    \delta(\varepsilon_{k})\delta(\varepsilon_{k'}) \label{eq_3}
\end{eqnarray}
%
%
where,
$g_{\textbf{k}\textbf{k}'}^{j}=[\hbar/2M\nu^{j}_{\textbf{k}'\textbf{k}}]^{1/2}\langle
\textbf{k}|\hat{\epsilon}^{\; j}_{\textbf{k}'\textbf{k}}\cdot\nabla
V|\textbf{k}'\rangle$ defines the matrix elements of electron-phonon interaction for electron
scattering from $\textbf{k}$ to $\textbf{k'}$ with a phonon of
frequency $\nu_{\textbf{k}-\textbf{k'}}^{j}$ (j is a branch index).
$M$ stands for the ion mass, $V$ is the crystal potential,
$\hat{\epsilon}^{\; j}_{\textbf{k}'\textbf{k}}$ is the polarisation
vector, and $N(0)=\sum_{k}\delta(\varepsilon_{k})$ represents the
single-spin electronic density of states at the Fermi surface. As
usual $\delta(x)$ denotes the Dirac's delta function evaluated at
$x'=0$. In addition, note that $|g_{\textbf{k}\textbf{k}'}^{j}|^{2}$ is inversely proportional to the number of
charge carriers contributed by each atom of the crystal to the
bosonic coupling mode. Therefore, an increase in the doping level,
which causes an increment in the hole concentration of the $CuO_{2}$
plane must be reflected in the coupling parameters as we will see
in section~\ref{Sec_2}. Moreover, recalling the
outstanding feature of the theory of metals, that
$|g_{\textbf{k}\textbf{k}'}^{j}|^{2}$ vanishes linearly with
$|\textbf{k}-\textbf{k}'|$ when $|\textbf{k}-\textbf{k}'| \ll
{k}_{F}$ \cite{Ashcroft76}, one would expect a {\em linear}
disappearance of the coupling effect that gives rise to the nodal
kink in the vicinity of the Fermi surface. On the other hand,
inspired by recent results on the ``universality'' of the nodal Fermi
velocity $v_{F<}$ (at low energies) in certain cuprates, a prominent
role of this quantity is also expected.

\subsection{The coupling parameters and thermodynamic properties}


Here, we recall that relevant dynamical information is contained in the
analytic continuation $G(k,\omega+i0^{+})$ to points just above the
real frequency axis, known as the ``retarded'' Green's function~\cite{Doniach98}. One
is therefore led to continue the electronic self-energy
$\Sigma(k,i\omega_{n})$ analytically by
$\Sigma(k,\omega+i0^{+})\equiv
\Sigma_{1}(k,\omega)+i\Sigma_{2}(k,\omega)$, where the bare electron band energy is determined by the poles of the Green's function $G(k,\omega+i0^{+})$ or the zeros of $G^{-1}(k,\omega+i0^{+})$. Assuming that a pole occurs near $\omega=0$, one gets
%
%
\begin{flushleft}
\begin{eqnarray}
G^{-1}(k,\omega+i0^{+})&&=\omega-\varepsilon_{k}-\Sigma_{1}(k,\omega)-i\Sigma_{2}(k,\omega) \nonumber \\&& 
\simeq\omega\left(1-\left.\frac{\partial\Sigma_{1}(k,\omega)}{\partial\omega}\right|_{\omega=0}\right)-\left[\varepsilon_{k}+\Sigma_{1}(k,0)\right]-i\Sigma_{2}(k,\omega)\, . \nonumber \\ \;\label{eq_4}%
\end{eqnarray}%
\end{flushleft}
Then, the pole of $G$ occurs at a frequency $\omega_{0}$ given by
$\omega_{0}=E_{k}-i/2\tau_{k}$, with the quasiparticle scattering time defined by 
$\tau_{k}^{-1}=-2\left(1-\partial_{\omega}\Sigma_{1}\right)^{-1}\Sigma_{2}(k,E_{k})$, and the electron dressed band energy $E_{k}$ by
%
%
\begin{equation}
E_{k}=(1-\partial_{\omega}\Sigma_{1})^{-1}\left[\varepsilon_{k}+\Sigma_{1}(k,0)\right]
\; .\label{eq_5}%
\end{equation}
%


Now, as a manifestation of the electron-phonon coupling interaction one can introduce the mass renormalisation of the electronic dispersion at the energy scale associated with the phonons. This may be technically defined by a mass-enhancement parameter $\lambda^{*}$ \cite{Grimvall81}, i.e., 
%
%
\begin{equation}
\lambda_{k}^{*}\equiv\left.-\partial_{\omega}\Sigma_{1}\right|_{\omega=0}
\; . \label{eq_6}%
\end{equation}


On the other hand, the strength of the electron-phonon interaction is commonly estimated from the so-called boson coupling parameter $\lambda$. This dimensionless parameter is commonly defined in terms of the electron-phonon spectral density as
%
%
\begin{equation}
\lambda\equiv2\int_{0}^{\infty}d\nu\,\frac{\alpha^{2}F(\nu )}{\nu}
\; , \label{eq_7}%
\end{equation}
and it will be eventually related to the superconducting transition temperature.


We want to emphasise that this quantity is not to be straightforwardly identified with the mass-enhancement
parameter $\lambda^{*}$ and a necessary distinction between them is essential for the overall description of the available ARPES data \cite{Ruiz09}. In fact, as it was argued in that paper, equality would just be warranted at low temperatures and within the Migdal-Eliashberg approximation for the phonon coupling.


\begin{figure}
\begin{center}
\includegraphics[width=.60\textwidth]{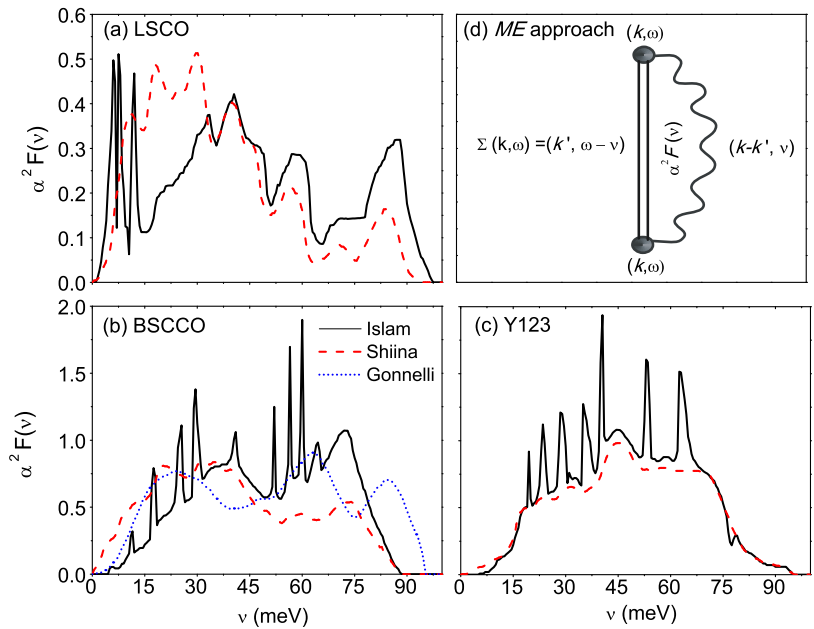}
\end{center}
\caption{\label{Fig_1}  Electron-phonon spectral density $\alpha^{2}\textit{F}(\nu)$ for (a) $La_{2-x}Sr_{x}CuO_{4}$ (LSCO), (b) $Bi_{2}Sr_{2}CaCu_{2}O_{8+x}$ (BSCCO), and (c) $YBa_{2}Cu_{3}O_{6+x}$ (YBCO), determined by different methods. The solid lines correspond to the method in Ref.~\cite{Islam00}, dashed lines to the method in Ref.~\cite{Shiina90}, and the dotted line in (b) corresponds to the method of Ref.~\cite{Gonnelli98}. In (d) we shown a schematic picture of the Migdal-Eliashberg approach from the lowest order Feynman diagram for the electron-phonon interaction. The wavy line represents the phonon Green's function $D$, the double solid line is the renormalised electron Green's function $G$, and the shaded circles represent the small electron-phonon vertex corrections neglected in the ME approach.}
\end{figure}
%


As one can easily appreciate, the spectral density function $\alpha^{2}F(\nu)$ plays a fundamental role in the electron-phonon coupling. Since the electron-phonon interaction manifests itself not only in the electronic properties but also in phononic properties via the density of phonon states $F(\nu)$, the investigation of phonon spectra may provide valuable information on the strength and character of electron-phonon coupling. The information about the phonon spectrum can be taken from the inversion of the Eliashberg equations in an analysis of tunnelling data~\cite{Allen82,Gonnelli98} or from inelastic neutron scattering data for superconductors~\cite{Islam00,Shiina90}. Along these lines, we have evaluated the electron spectral densities for LSCO, BSCCO, and YBCO in the flat model of Ref.~\cite{Shiina90} and also by solving the isotropic Eliashberg equations on the Matsubara frequencies. Strictly speaking, the structure of these spectral densities is restricted to the isotropic nodal direction under the assumption that $\alpha^{2}F(\nu)=G(\nu)\times C$, with C an adjustable constant and $G(\nu)$ the generalised phonon density of states extracted from the inelastic neutron scattering experiments~\cite{Shiina90}. Our results are shown in Fig. 1 (Shiina). In addition, other reproducible methods to calculate the electron-phonon spectral density have been taken into consideration, avoiding the intrinsic complexity in the evaluation of the matrix elements that determine the spectral density $\alpha^{2}F(\nu)$ from ab initio calculations. Specifically, we mean the simple method by Islam \& Islam~\cite{Islam00} for all samples, and the method by Gonnelli et al.~\cite{Gonnelli98} in the particular case of BSCCO. The method of Refs.~\cite{Islam00} \& \cite{Shiina90} is based on the INS experimental data by Renker \textit{et al.} \cite{Renker87,Renker88,Renker89}. On the other hand the method within Ref.~\cite{Gonnelli98} is based on the tunnelling data reported in the Refs.~\cite{Gonnelli97,Ummarino97}. 


Below, we consider the possibility that the electron-phonon spectral densities presented above (Fig.~\ref{Fig_1}) allow to explain the high-$T_{c}$ values, and the zero temperature gap observed in experiments. 


\begin{figure}
\begin{center}
\includegraphics[width=.6\textwidth]{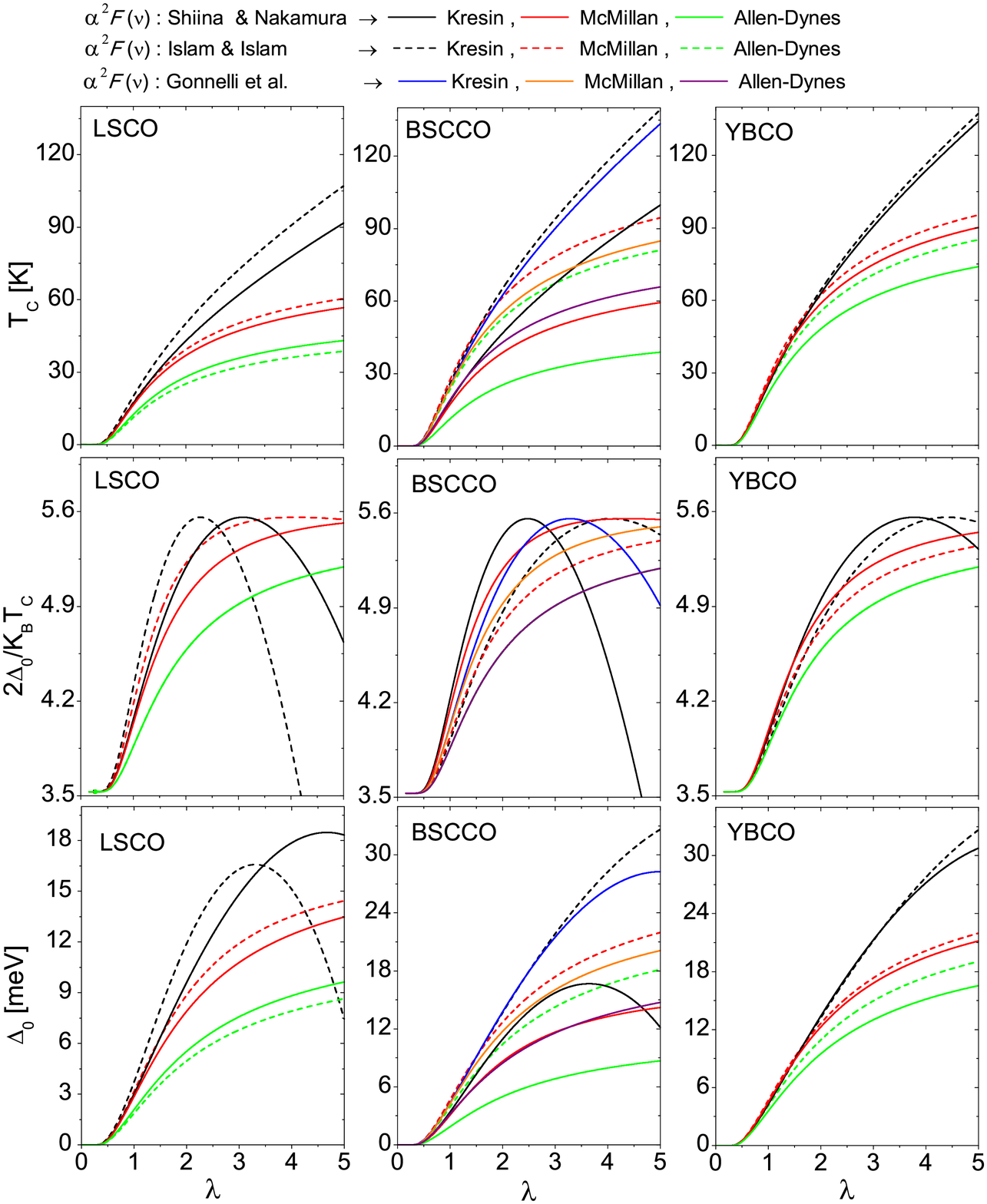}
\end{center}
\caption{\label{Fig_2}  Plots of the
critical temperatures $T_{c}$ (top), the ratio gap $2\Delta_{0}/k_{B}T_{c}$ (middle) and the gap $\Delta_{0}$ (bottom) for LSCO, BSCCO and YBCO,  all of them represented as functions of the electron-phonon coupling parameter $\lambda$, and corresponding to the spectral densities of Shiina \& Nakamura~\cite{Shiina90} and Islam \& Islam~\cite{Islam00}. In addition the spectral density of Gonnelli et al.~\cite{Gonnelli98} has been considered in the BSCCO case (Fig.~\ref{Fig_1}). We have used three different approximations, i.e.: McMillan's formula~\cite{McMillan68}, the Allen-Dynes formula~\cite{Allen75}, and Kresin's formula~\cite{Kresin87}.}
\end{figure}


In Fig.~\ref{Fig_2} we show our results for $T_{c}$, the ratio gap $2\Delta_{0}/k_{B}T_{c}$ and the zero temperature gap $\Delta_{0}$, all of them based on three different approximations:\\
(1) The celebrated McMillan's equation \cite{McMillan68}, 
%
%
\begin{eqnarray}
T_{c}=\frac{\omega_{1}}{1.2}exp\left[-1.04\frac{1+\lambda}{\lambda-\mu^{*}(1+0.62\lambda)}\right]
\;\label{eq_8}
\end{eqnarray}
with the phonon characteristic energy defined by $\omega_{\rm 1}=2S/\lambda$, S being the area under the spectral density $\alpha^{2}F(\nu)$, and $\mu^{*}$ the so-called Coulomb pseudopotential\\
(2) The Allen and Dynes formula \cite{Allen75}, which is obtained by replacing $\omega_{1}$ in Eq.~(\ref{eq_8}) by
%
%
\begin{eqnarray}
\omega_{\rm log}\equiv exp\left\{\frac{2}{\lambda}\int_{0}^{\infty}d\nu\frac{\alpha^{2}F(\nu) \ln(\nu)}{\nu}\right\}
\; .\label{eq_9}
\end{eqnarray}
In the systems analysed here,  we get $\omega_{\rm log}^{LSCO}\simeq 16\, meV$, $\omega_{1}^{LSCO}\simeq 25\, meV$, $\omega_{\rm log}^{BSCCO}\simeq 33\, meV$, $\omega_{1}^{LSCO}\simeq 39\, meV$, $\omega_{\rm log}^{YBCO}\simeq 35\, meV$, $\omega_{1}^{LSCO}\simeq 39\, meV$.\\
(3) Finally, the less conventional Kresin's formula \cite{Kresin87},
%
%
\begin{eqnarray}
T_{c}=0.25\; \varpi \; exp\left(\frac{2}{\lambda_{eff}}-1\right)^{-1/2}
\; ,\label{eq_10}
\end{eqnarray}
where $\varpi=[(2/\lambda)\int_{0}^{\infty}\nu\alpha^{2}F(\nu)d\nu]^{1/2}$ and $\lambda_{eff}=(\lambda-\mu^{*})[1+2\mu^{*} + (3/2) \lambda \mu^{*} exp(-0.28\lambda)]$.


As recalled before \cite{Ruiz09}, a satisfactory perturbation theory for the Coulomb interaction does not exist. Then, a phenomenological point of view is usually adopted, considering that the
superconducting components of the Coulomb self-energy turn out to
have only a small effect over the bare energy of the quasiparticles. In this sense, in a first instance, we considered a typical value of the Coulomb's pseudopotential, $\mu^{*}=0.13$ \cite{Ruiz09}. In this paper, the Coulomb effects and their influence on the electron-phonon interaction picture will be thoroughly discussed. 

As regards the ratio gap, this quantity has been calculated by using the common formula, $2\Delta_{0}/k_{B}T_{c}=3.53~[1+12.5~(T_{c}/\omega_{\rm log})^{2}~ \rm ln(\omega_{\rm log}/2T_{c})]$.


From Fig.~\ref{Fig_2} several conclusions can be drawn about the relevance of the phonon-mechanism~\cite{Islam00,Shiina90}. Apparently, from the theoretical point of view the high values of the critical temperatures strongly depend on the approximation used to calculate them, and even in some cases on the inversion method used for obtain the spectral density. In this sense, the electron-phonon coupling parameter $\lambda$ has been usually considered a fit parameter used for obtaining the correct experimental critical temperatures, and then used to conclude strong or weak influence of the electron-phonon interaction on the superconductivity properties. Nevertheless, the strong dependence of the derived parameters on the particular formulae considered, does not favour this procedure. Then, our position here is determine the electron-phonon coupling parameter from experimental available spectroscopies as the energy dispersion curves obtained in ARPES experiments and compare it with the $T_c$ value calculated with different approximations an different spectral functions. Outstandingly, the clearest point is to check if, regardless the approach used for obtaining the spectral function, it leads or not to correct predictions for the observed values of $T_{c}$ and $\Delta_{0}$. 


We should comment that other choices of the spectral density in which the interaction mechanism has a magnetic origin are also possible. Nevertheless, in this picture, one must accept that the magnetic mode can be described by the same Feynman diagram as the phonon mode (Fig.~\ref{Fig_1}d). This hypothesis was first proposed by Berk and Schrieffer~\cite{Berk66} and recently invoked by Dahm \textit{et al.}~\cite{Dahm09}.  To support our ``phononic choice'' we want to recall that the absence of the magnetic-resonance mode in LSCO \cite{Zhou05}, BSCCO over-doped (x=0.23) \cite{Hwang04} and its appearance only below $T_{c}$ in some cuprates (e.g., Bi2201 \cite{Sato03}) are not consistent with the idea that the kink dispersion has a magnetic origin. On the other hand, recent studies on electron doped systems \cite{Park08,Graf07,Wilson06,Zhao07} have shown that the intensity of the  magnetic resonance mode is seemingly weak in comparison with the phonon mode to be considered as the cause of the strong electron energy dispersion measured by ARPES. However, this issue will be reconsidered in our final discussion.

\section{Nodal ARPES spectra}
\label{Sec_2}


It is well known that angle-resolved photoemission spectroscopy (ARPES) has become a powerful technique for evaluating the quasiparticle self-energies $\Sigma(\textbf{k},\omega)$ in condensed matter, despite the complexities in understanding the photoemission processes involved (refer to our discussion in Ref.\cite{Ruiz09} and references therein). On the other hand, within the electron-phonon coupling scenario and starting from Eq.(\ref{eq_2}) within the ME approach, the electronic self-energy can be obtained from the complex expression
%
%
\begin{equation}
\Sigma(\textbf{k},i\omega)=\int_{0}^{\infty}d\nu\alpha^{2}F(\textbf{k},\nu)\left\{-2\pi
i\left[N(\nu)+\frac{1}{2}\right]+\Upsilon(\nu,\omega,T)\right\}\, ,\label{eq_11} 
\end{equation}
with the function $\Upsilon(\nu,\omega,T)$ defined by
%
%
\begin{equation}
\Upsilon(\nu,\omega,T)=\psi\left(\frac{1}{2}+i\frac{\nu-\omega}{2\pi
T}\right)-\psi\left(\frac{1}{2}-i\frac{\nu+\omega}{2\pi T}\right)\, .\label{eq_12} 
\end{equation}
%
%
%
Here $\psi(z)$ are the so-called digamma functions. Now, recall that the bare band energy $\varepsilon_{\textbf{k}}$ relates to the dressed band energy $E_{\textbf{k}}$ by $E_{\textbf{k}}=\varepsilon_{\textbf{k}}+Re\Sigma(E_{k})$. Whereas the direct extraction of the self-energy from experiments appears to be troublesome because the underlying band structure of the bare electrons is \textit{a priori} unknown, a theoretical determination of the bare band structure and its relation to the full energy renormalisation effects observed in the experiments seems much more attractive. In this sense, the nodal ARPES spectra discussed below are of great importance to check the validity of quasiparticle concept and understanding the nature of the interactions involved.


As a description of the ARPES technique and related photoemission spectroscopies is beyond the scope of this paper, we consider a pertinent suggestion for the readers the review by A. Damascelli, Z.-X. Shen and Z. Hussain [Ref.~\cite{Damascelli03}], and the works contained in the most recent books of Stefan H\"{u}fner [Ref.~\cite{Hufner07}], X. J. Zhou \textit{et al.} [Ref.~\cite{Zhou07}] and by J. R. Schrieffer and J. S. Brooks [Ref.~\cite{Schrieffer07}].

\begin{figure*}
\begin{center}
\includegraphics[width=1.0\textwidth]{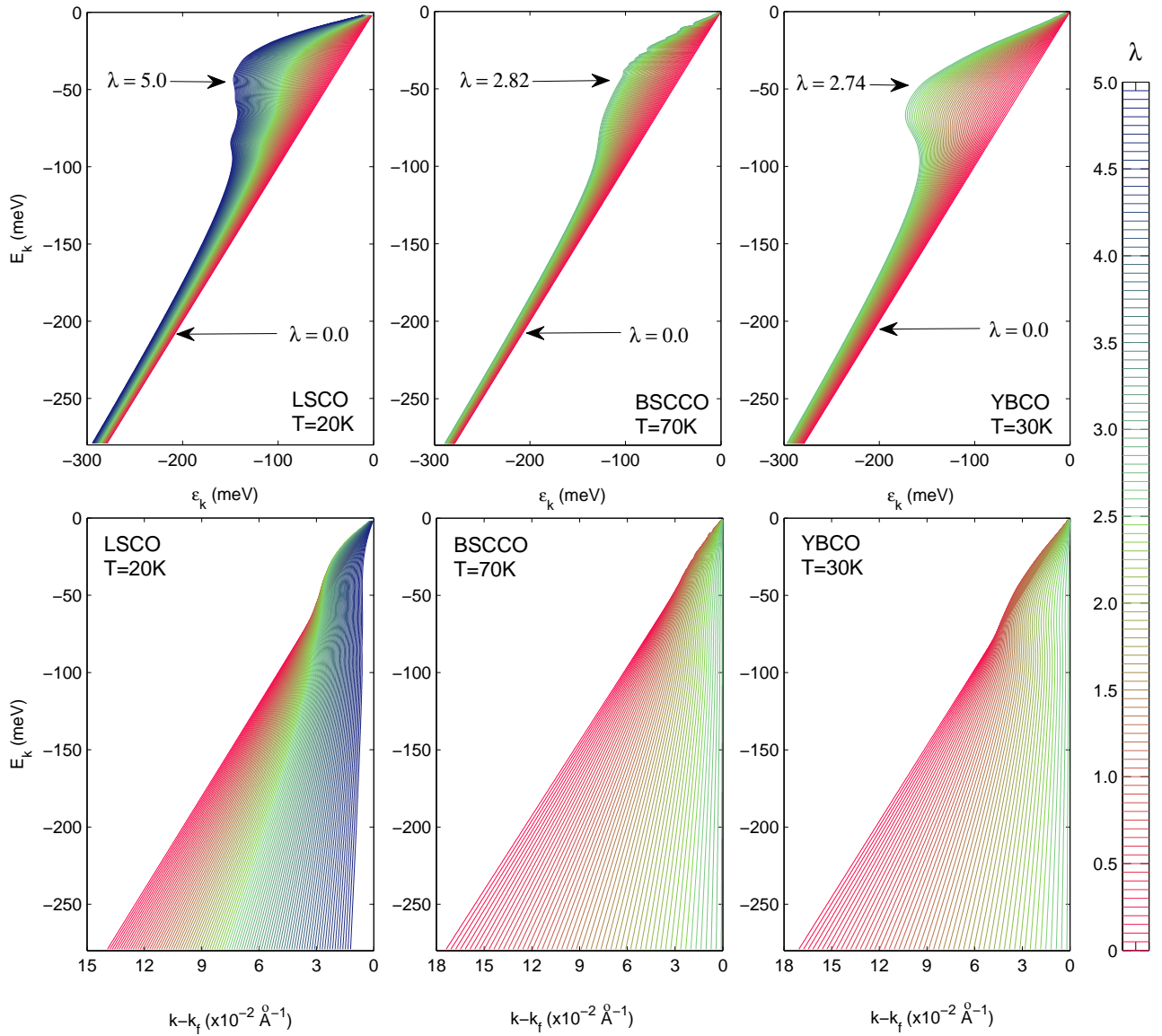} 
\end{center}
\caption{\label{Fig_3}  Top: The renormalised energy $E_{k}$ as a function of the bare band energy $\varepsilon_{k}$ assuming that the nodal dispersion is a consequence of the electron-phonon coupling. All curves have been obtained using Eq.~(\ref{eq_11}) and the EPI spectral density $\alpha^2\emph{F}(\omega)$ calculated by the more rigorous  method of Ref. \cite{Shiina90} (Fig.~\ref{Fig_1}) in LSCO (left), BSCCO (centre) and YBCO (right). Very similar results are found under the use of any of the densities shown in the Fig.~\ref{Fig_1}. The curves are labelled according to the e-ph coupling parameter ``$\lambda$'' and range between the bare dispersion $\lambda=0.0$ (straight line) and about $\lambda=4.0$ for LSCO (at T=20K), $\lambda=2.82$ for BSCCO (at T=70K) and $\lambda=2.74$ for YBCO (at T=30K). Bottom: Simulation of the corresponding renormalised energies $E_{k}$ as a function of the momentum $k-k_{F}$ along the nodal direction from our own model (Eq.(\ref{eq_13})) with $\delta_{LSCO}=0.185$, $\delta_{BSCCO}=0.354$, and $\delta_{YBCO}=0.365$.}
\end{figure*}

The purpose of this section is to go through the prominent results obtained from ARPES measurements on the (0,0)-($\pi,\pi$) direction in the Brillouin zone (nodal direction) avoiding the complications of the superconducting gap. We want to emphasize that in the nodal direction the \emph{d}-wave superconducting gap in cuprates is zero and the interactions involving a low-energy excitation appear as a well-defined slope change in the electronic energy-momentum dispersion in a similar energy scale ($E-E_{F}\sim40-80 meV$ or \textit{kink}~\cite{Lanzara01,Zhou03,Zhou05,Kordyuk06,Xiao07,Borisenko06} (see Fig.~\ref{Fig_3}). On the other hand, the antinodal direction denotes the $(\pi,0)$ region where the anisotropic character of the \emph{d}-wave superconducting gap seems to be unavoidable~\cite{Zhou02,Xiao07,Takahashi07,Gweon04,Douglas07}. In fact, any theoretical assumption to include the \emph{d}-wave character of the gap function could be clouding the real influence of the phonon-coupling modes or even any other mechanism, strong or weak, it involved in the superconducting pairing properties. In this sense, is not queer that in the antinodal direction the kink-topology is not universal in comparison with the called universal nodal Fermi velocity~\cite{Zhou03}. Special attention will be payed on the electron-phonon interaction in HTSC-cuprate compounds along the lines of the previous discussion.


Let us to mention that the energy distribution- and momentum distribution-curves (EDC - MDC) are the two most popular ways for analysing photoemission data. The dichotomy between the MDC- and EDC-derived bands from the same data raises critical questions about its origin and also about which one represents the intrinsic band structure. In this sense, we want to recall that at the larger bandwidth along the nodal direction, the MDC method can be reliably used to extract high quality data of dispersion in searching for fine structure. It has also been shown theoretically that this approach is reasonable in spite of the momentum-dependent coupling if we are only interested in identifying the mode energies \cite{Devereaux04}. In a typical Fermi liquid picture, the MDC- and EDC-derived dispersions are identical. Moreover, in an electron-boson coupling system the lower and higher energy regions of the MDC- and EDC-derived dispersions are still consistent, except right over the kink region \cite{Zhang08}. Below, the good agreement between our theoretical data and the experimental facts supports this general picture.


Considering that the kink effect is a common feature of the HTSC cuprates, i.e., it is observable regardless of the doping level or the temperature at which the measurement is performed, the only possible scenario seems to be the coupling of quasiparticle with phonons. The key issue seems to be the existence of an energy scale (in the range of 40-80 meV).  As in previous work, our position in this paper has been to determine the evolution of the bare electron band energy $\varepsilon_{k}$ through the systematic evaluation of the dressed band energy $E_{k}$ as a function of the electron-phonon coupling $\lambda$ in several HTSC cuprates, i.e., LSCO, BSCCO and YBCO (Fig.~\ref{Fig_3}). To be specific, we propose to generalise use  of the  equation $E_{k}=\varepsilon_{k}+Re\{\Sigma(E_{k})\}$ from which one can consider that $\varepsilon_k$ implicitly depends on $E_k$ through the boson coupling parameter $\lambda$. We recall that $\varepsilon_{k}$ is not directly available from the experiments. Instead, the electron momentum dispersion curve $E_{k}(k-k_{F})$ may be measured from the ARPES experiments. This fact, along with an ansatz for the ARPES ``bare'' dispersion will allow to obtain $\lambda$ as a unique constrained parameter that better fits the observed kink topology in a particular sample \cite{Reznik08,Dahm09}. Thus, based on an interpolation scheme between the numerical behaviour of the dressed energy band and the experimental data, one can introduce a universal dispersion relation \cite{Ruiz09} which allows to reproduce in a quite general scheme the nodal dispersions close to the Fermi level in LSCO (Fig.~\ref{Fig_4}), BSCCO (Fig.~\ref{Fig_6}), and YBCO (Fig.~\ref{Fig_7}) samples. 


As discussed in Ref.\cite{Ruiz09}, to a first approach, our relation is defined by
%
%
\begin{equation}
{k}-{k}_{F}\approx\frac{\varepsilon_{k}}{v_{F_{<}}}\left(1-\delta\lambda\right)\, ,\label{eq_13}
\end{equation}
%
with $v_{F_{<}}$ the Fermi velocity at low-energies. Recall that, contrary to the behaviour at high energy values, this limit of the Fermi velocity is rather independent of the chemical structure and doping levels and can be obtained as the slope of the lower part of the momentum dispersion curve \cite{Zhou03}. Note that then, $\delta$ is the only ``\textit{free}'' parameter required for incorporating the specific renormalisation for a given superconductor. Let us emphasise once again that the correct mass-enhancement parameter defined in Eq.~\ref{eq_6} cannot be directly identified with the electron-phonon coupling parameter $\lambda$, as well as the complete electron self-energy $\Sigma_{1}$ is not to be straightforwardly identified with the electron-phonon self energy $\Sigma_{ep}$ (though it is an important part of it). As a first approximation to this fact, recall that in Eq.~(\ref{eq_13}) the involved quantities can be assumed to be not far from their values at the Fermi level and consequently, one is enabled to replace $\varepsilon_{k}$ by
%
%
\begin{equation}
E_{k}-\Sigma_{1}(E_k)=\left({k}-{k}_{F}\right)v_{F_{<}}\left(1-\delta\lambda\right)^{-1}\,
.\label{eq_14}
\end{equation}
As it was shown in Ref.~\cite{Ruiz09}, starting from this relation and Eq.~(\ref{eq_6}) one can show that the parameter $\delta$ is to be physically defined by $\delta\equiv(\lambda^{*}/\lambda)/(1+\lambda^{*})$; i.e., $(1-\delta\lambda)^{-1}=1+\lambda^{*}$. Recall that the proof relies on the fact that $\lambda^{*}$ is defined from the electronic group velocity $\textbf{\textit{v}}_{\textbf{k}}=(1/\hbar)\partial E/\partial\textbf{k}$, which changes by a factor of $1/(1+\lambda^{*})$ close to the Fermi level~\cite{Ashcroft76}. This allows to write Eq.~(\ref{eq_12}) in the alternative way
%
%
\begin{equation}
{k}-{k}_{F}=\frac{\varepsilon_{k}}{v_{F_{<}}}\frac{1}{1+\lambda^{*}} \, .\label{eq_15}
\end{equation}
Comparing both equations, to the lowest order, the dimensionless parameter $\delta$ is basically the ratio between the defined mass-enhancement and phonon-coupling parameters
$\delta\approx\lambda^{*}/\lambda$. Remarkably, the most general Eq.~(\ref{eq_15}) allows to consider higher-orders in the electron-phonon coupling from the physical meaning of $ \delta$, or even consider additional mechanisms from the physical meaning of $\lambda^{*}$.

In this work, we have extended previous studies to the consideration of different spectral densities, and also analysed the implications of different schemes for obtaining $T_c$. This will allow to check the relevance of the phonon mechanism and the relevance of Eq.(\ref{eq_13}) for obtaining the coupling parameter $\lambda$.
 The main facts that arise from this analysis are developed in following subsections, and summarised in figures~\ref{Fig_4}~-~\ref{Fig_9}.


\subsection{Results in LSCO}


In Fig.~\ref{Fig_3} we show the singular features involved within the bare electron band energy curves [$E_{k}(\varepsilon_{k})$] and the renormalised dressed electron band energies [$E_{k}(k-k_{F})$]. One can observe a structure in the renormalised quasiparticle energy $E_{k}$ ranging up to roughly 80 meV. In fact, the bands rapidly approach $E_{F}$ from high-binding energy and suddenly bend at 40-80 meV, showing the kink topology described in the ARPES experiments.


\begin{figure*}
\begin{center}
\includegraphics[width=1.0\textwidth]{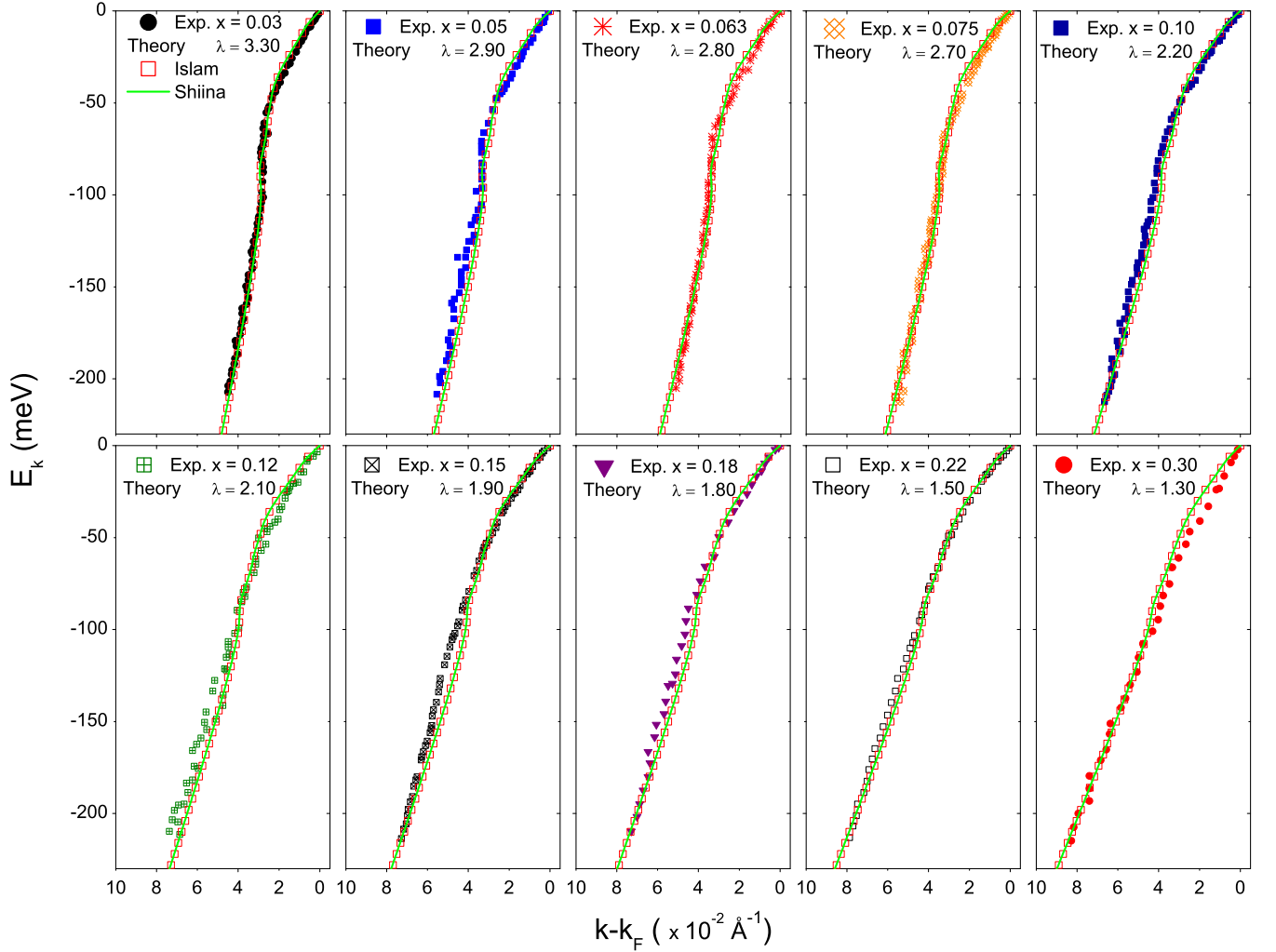} 
\end{center}
\caption{\label{Fig_4}  The renormalised electron quasiparticle energy dispersion $E_{k}$ as a function of the momentum $k-k_{F}$ for several samples of $\texttt{La}_{2-x}\texttt{Sr}_{x}\texttt{CuO}_{4}$, measured along the (0,0)-($\pi$,$\pi$) nodal direction at a temperature of 20 K. The doping level $x$ ranges between 0.03 (top left) up to 0.30 (bottom right). The experimental data (symbols) are taken from Ref.~\cite{Zhou03} and the theoretical curves have been obtained using the spectral densities of Islam \& Islam [Ref.~\cite{Islam00}] (open squares), and  Shiina \& Nakamura (solid lines)\cite{Shiina90}.]}
\end{figure*}


For quantitative purposes, here, we have considered $v_{F_{<}}=2 eV\cdot${\AA} as related to the experimental data of Refs. \cite{Zhou03,Zhou07,Lanzara01,Zhou05}. In Fig.~\ref{Fig_4} we show the results found in $La_{2-x}Sr_{x}CuO_{4}$ covering the doping range ($0<x\leq0.3$). Remarkably, in this material, the hole concentration in the $\texttt{CuO}_{2}$ plane is well controlled by the \texttt{Sr} content $x$ and a small oxygen non-stoichiometry. Within this range, the physical properties span over the insulating, superconducting, and over-doped non-superconducting metal behaviour. Superconducting transition temperatures $T_{c}$ in the interval of 30-40K have been observed by Bednorz and M\"{u}ller~\cite{Bednorz86} and others~\cite{Uchida87,Dietrich87}. On the other hand, the best fit of the whole set of experimental data has been obtained for $\delta=0.185$ (Fig.~\ref{Fig_4}).


\begin{figure}
\begin{center}
\includegraphics[height=.35\textwidth]{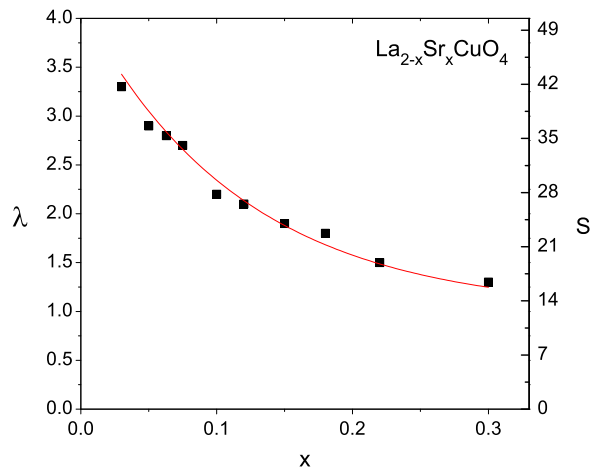} 
\end{center}
\caption{\label{Fig_5} Evolution of the e-ph coupling parameter $\lambda$ as a function of the dopant content in $La_{2-x}Sr_{x}Cu_{4}$ samples. The $\lambda$-values have been obtained from our best fit with the nodal kink dispersion (black squares) shown in fig.~\ref{Fig_4}. Correspondingly, the evolution of the area $S$ as a function of the dopant content is also shown (right scale).}
\end{figure}


In fig.~\ref{Fig_5} the evolution of the parameter $\lambda$ as a function of the doping level is shown. Taking advantage of the widespread availability of experimental information in LSCO, one can fit the data to the simple expression $\lambda=2\tilde{\omega}exp(-\frac{\tilde{\omega}}{\delta} x)+1$, within a precision factor around of 95\%. $\tilde{\omega}$ is the ratio between the phonon characteristic energies introduced by McMillan \cite{McMillan68}, $\omega_{\rm 1}=(2/\lambda)\int_{0}^{\infty}\alpha^{2}F(\nu)d\nu\equiv (2/\lambda)S$, and Allen and Dynes \cite{Allen75}, $\omega_{\rm log}\equiv exp\{(2/\lambda)\int_{0}^{\infty}\ln(\nu)[{\alpha^{2}F(\nu)}/{\nu}]d\nu\}$, i.e.,  $\tilde{\omega}$=$\omega_{1}/\omega_{\rm log}$. We get $\omega_{\rm log}^{LSCO}\simeq 16.1455\, meV$ and $\omega_{1}^{LSCO}\simeq 25.2627\, meV$. We want to clarify that, although this equation can be just considered as a useful relation between the physical and chemical properties of LSCO, the shaping of other HTSC by similar expressions cannot be guaranteed. Regarding the critical temperatures, it must be emphasised that for LSCO, the $\lambda$-values obtained from our model \cite{Ruiz09} allow to explain the observed superconducting temperatures $T_{c}$ \cite{Bednorz86,Uchida87,Dietrich87}. This can be done either by using the McMillan formula \cite{McMillan68} or in a more general context, by solving the Eliashberg equations \cite{Shiina90}. Nevertheless, we want to emphasise on how these facts have to be interpreted recalling that several discrepancies could be emerge depending on the formula used for obtaining $T_{c}$. From the previous section we have concluded that the critical temperature $T_{c}$ should not be considered as a fit parameter for adjusting the theory, i.e.: one should not predict $\lambda$ from the approximate $T_{c}$ formulas and then use it for calculating the electron self-energy. Outstandingly, this attractive idea has led to unfortunate underestimates of the phonon contribution to the photoemission kink in HTSC \cite{Giustino08,Heid08}. On the other hand, to our knowledge, the most suitable way for determining the influence of an interaction mechanism in the pair formation for HTSC is (i) evaluate the strength of the boson coupling mode from the electron renormalisation effects and then (ii) solve the Eliashberg equations for the superconducting $T_{c}$. In addition, a semiempirical approach has been introduced in this paper by using the celebrated equations for $T_{c}(\lambda)$ referred above. From such analysis, we conclude that the consideration of the electron-phonon interaction in LSCO strongly suggests that the high $T_{c}$ values can be caused by conventional electron-phonon coupling, in agreement with the conclusion of Weber \cite{Weber87}. In that case they obtained $\lambda=2-2.5$ in the range $0.2>x>0.1$ within the framework of the nonorthogonal-tight-binding theory of lattice dynamics, based on the energy band results of Mattheiss~\cite{Mattheiss87}, and remarkably corresponding to the range of the EPI coupling parameter  $\lambda$ predicted in our model. Moreover, within our model the quasiparticle parameter ``$\lambda$'' calculated from the gradient of the DFT self-energy is basically to be identified with the mass-enhancement parameter $\lambda^{*}$. Thus, although the density-functional theory gives a correct ground-state energy, the bands do not necessarily fit the quasi-particle band structure used to describe low-lying excitations in agreement with Ref. \cite{Reznik08}, and the experimental quasiparticle parameter $\lambda_{expt}$ matches the electron phonon coupling $\lambda$ involved in the standard Migdal formalism for analysing the experiments~\cite{Ruiz09}. On the other hand, it must be emphasised that in the case of Ref.~\cite{Weber87}, the $T_c$ values were also reproduced. Moderate discrepancies between the predictions of our phenomenological model and the analysis of Refs.~\cite{Weber87} \& \cite{Weber88} may be ascribed to some uncertainty in the experimental spectral densities. Remarkably the high $T_{c}$ values observed in LSCO have been obtained from the McMillan equation, or in a most general way by solving the Eliashberg equations according to Ref.~\cite{Shiina90}.


\subsection{Results in BSCCO: analysis of ARPES data}


\begin{figure}
\begin{center}
\includegraphics[width=.60\textwidth]{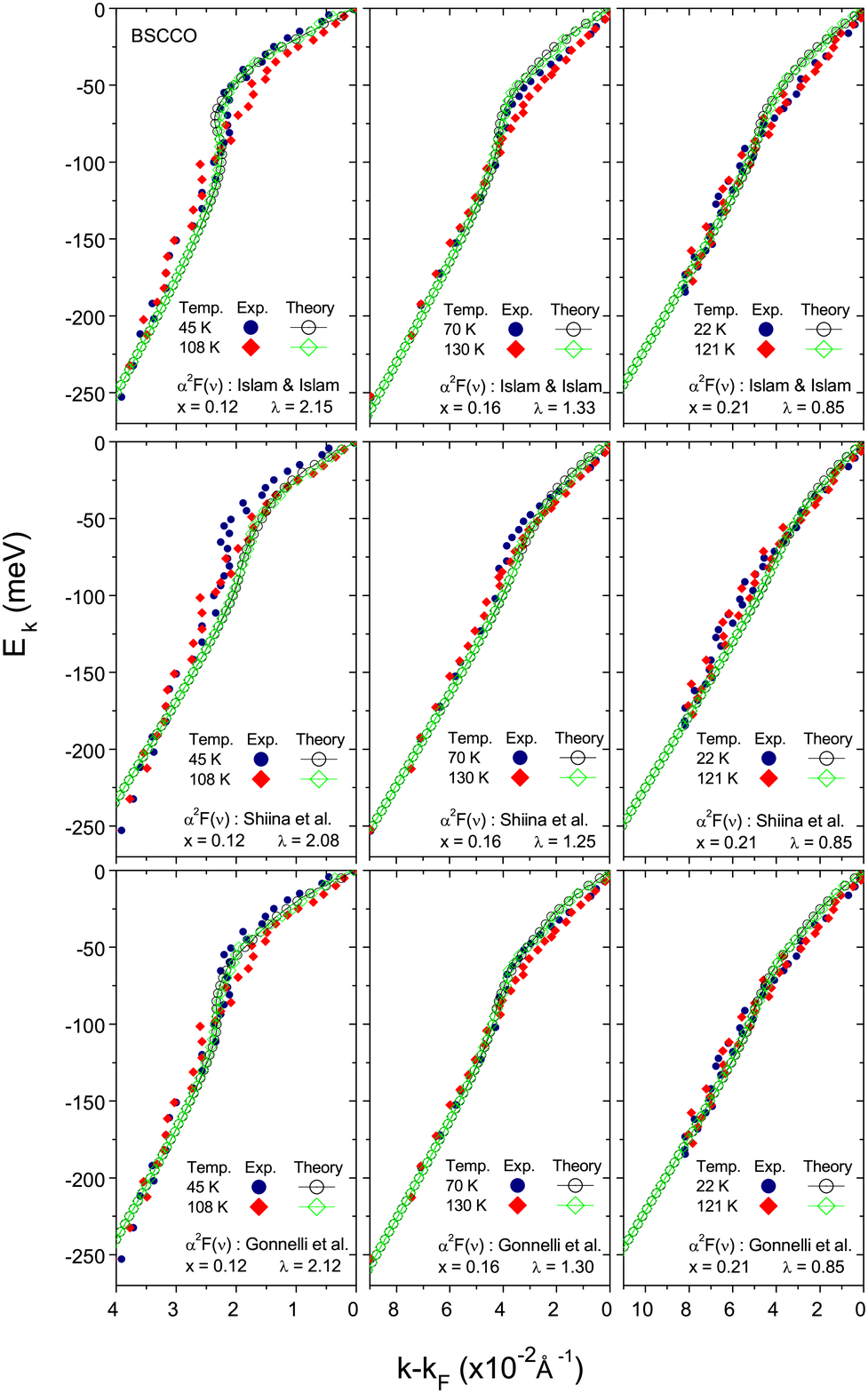}
\caption{\label{Fig_6} Same as in Fig.~\ref{Fig_4} but in samples of $\texttt{Bi}_{2}\texttt{Sr}_{2}\texttt{CaCu}_{2}\texttt{O}_{8+x}$, all measured along the (0,0)-($\pi$,$\pi$) nodal direction. The experimental data for the normal state (full diamond) and superconducting state (full circle) both have been taken from the  Ref.~\cite{Johnson01}. The theoretical curves (lines -diamond, or -circles) have been obtained from our dispersion relation (Eq.~\ref{eq_13}) according to the best fit values for e-ph coupling parameter $\lambda$ using the spectral densities of Islam \& Islam [Ref.~\cite{Islam00}] (top), Shiina \& Nakamura [Ref.~\cite{Shiina90}] (middle), and Gonnelli et al. [Ref.~\cite{Gonnelli97}] (bottom). The different plots correspond to the doping levels of BSCCO: under-doped ``x=0.12'' (left), optimally doped ``x=0.16'' (centre), and over-doped ``x=0.21'' (right) respectively.}
\end{center}
\end{figure}

In Fig.~\ref{Fig_6} we display the results found for the renormalised electron quasiparticle dispersion $E_{k}$ as a function of the momentum $k-k_{F}$ for several samples of the double layer $Bi_{2}Sr_{2}CaCu_{2}O_{8+x}$ measured along the (0,0)-($\pi$,$\pi$) nodal direction. BSCCO curves were predicted by our interpolation method and different inversion methods to determine the electron-phonon spectral densities (see fig.~\ref{Fig_1}), for the under-doped (UD70) ``$x=0.12$, $T_{c}\approx 70K$'', optimally doped (0PD90) ``$x=0.16$, $T_{c}\approx 90K$'', and over-doped (OVD58) ``$x=0.21$, $T_{c}\approx 58K$'' samples. The experimental data were taken from the work by Johnson \emph{et al.}~\cite{Johnson01} with $v_{F_{<}}=1.6eV\cdot${\AA} as a value consistent with the experimental results of Refs.~\cite{Lanzara01,Zhou02,Kordyuk06,Johnson01}. In a first approach, the best fit with experimental data has been found for $\delta=0.354$ within the same approximations done in the previous section, i.e., for determining the critical temperatures we have assumed $\mu^{*}=0.13$. On the other hand, in this work we show that regardless the method used for obtaining the electron-phonon spectral density (fig.~\ref{Fig_1}) the same conclusions are obtained (fig.~\ref{Fig_6}). Recall that, similarly to the case of LSCO, our analysis fits well the ``$\lambda$'' values predicted by other authors and from very different models~\cite{Kordyuk06,Gonnelli98,Shiina90} (see table~2 in Ref.\cite{Ruiz09}). However, as regards to the critical temperatures, only for the BSCCO UD70 sample the $\lambda$ values obtained can be consistent with the experimental facts, i. e., from the spectral density of Islam \& Islam the critical temperatures obtained are: $T_{c}^{Kresin}(\lambda=2.15)=70.0 K$ and  $T_{c}^{McMillan}(\lambda=2.15)=64.81 K$. From the spectral density by Gonnelli et al.: $T_{c}^{Kresin}(\lambda=2.15)=67.06 K$ and $T_{c}^{McMillan}(\lambda=2.15)=58.19 K$. However, we must mention that a strong reduction of $T_{c}$ ($\sim 40\%$) is found from the method by Shiina \& Nakamura~\cite{Shiina90} and the Allen \& Dynes formula~\cite{Allen75}. Perhaps, this fact could be considered as a first signal about the need of considering additional perturbation mechanisms previously not involved in the complete solution of the matrix elements for the Eliashberg equations, or perhaps and even too, this fact reveals that the flat model used in the Ref.~\cite{Shiina90} is not consistent with this experimental facts. This will be further discussed in a forthcoming section (Sec.\ref{sec_ephBY}).


\subsection{Results in YBCO: analysis of ARPES data}

\begin{figure}
\begin{center}
\includegraphics[width=.48\textwidth]{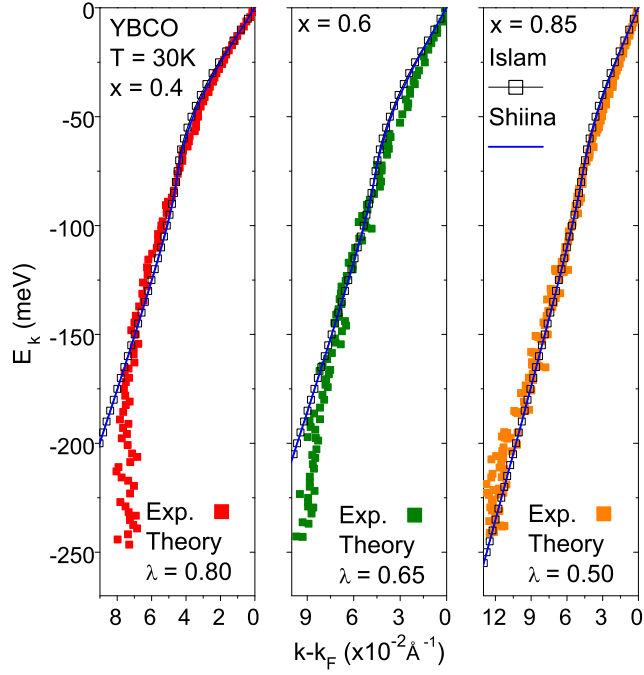}
\caption{\label{Fig_7} Same as in Fig.~\ref{Fig_4} but in samples of  $\texttt{Y}\texttt{Ba}_{2}\texttt{Cu}_{3}\texttt{O}_{6+x}$, all measured along the (0,0)-($\pi$,$\pi$) nodal direction. The experimental data have been taken from Ref.~\cite{Borisenko06}. The theoretical curves have been obtained from our dispersion relation [Eq.~(\ref{eq_13})] according to the best fit values for e-ph coupling parameter $\lambda$ using the spectral densities of Islam \& Islam [Ref.~\cite{Islam00}] (open squares), and Shiina \& Nakamura [Ref.~\cite{Shiina90}] (solid line). The different plots correspond to the doping levels:``x=0.4'' (under-doped), ``x=0.6'', and ``x=0.85'' (over-doped) respectively.}
\end{center}
\end{figure}

Regarding the YBCO family, that is characterised by double $CuO_{2}$ planes sandwiched between single $CuO$ chains a similar analysis has been done, but remarkable difficulties have been found. 


In Fig.~\ref{Fig_7} we display the results found for renormalised electron quasiparticle dispersion $E_{k}$ as a function of the momentum $k-k_{F}$ for several samples of $YBa_{2}Cu_{3}O_{6+x}$ measured along the (0,0)-($\pi$,$\pi$) nodal direction. The data were collected from our interpolation method and different inversion methods to determine the electron-phonon spectral densities (see fig.\ref{Fig_1}), in the under-doped ``$x=0.4$, $T_{c}\approx 35K$''(UD35), ``$x=0.6$, $T_{c}\approx 61K$''(UD61), and over-doped ``$x=0.85$, $T_{c}\approx 90K$''(OVD90) samples. The experimental data were taken from the work by Borisenko \emph{et al.}~\cite{Borisenko06} with $v_{F_{<}}=1.63eV\cdot${\AA} as a value consistent with the experimental results reported by those authors. In the first approach, the best fit with experimental data has been found for $\delta=0.365$~\cite{Ruiz09}. In this work we show that independently of the method used for obtaining the electron-phonon spectral density (Fig.~\ref{Fig_1}) the same conclusions are obtained (fig.~\ref{Fig_7}).  Recall that, similarly to the LSCO and BSCCO cases, our analysis fits well the ``$\lambda$'' values predicted by other authors and from very different models~\cite{Shiina90,Weber88,Heid08} (see table~3 in \cite{Ruiz09}). On the other hand, as regards $T_{c}$, we have previously concluded that the electron-phonon contribution is very weak considering the low critical temperatures obtained by different approaches notwithstanding, the kink effect was entirely reproduced~\cite{Ruiz09}. A discussion on the mismatch between the electron-phonon coupling parameters and the superconducting properties in YBCO is put forward in the following section.


\subsection{Electron-phonon coupling and superconductivity in BSCCO and YBCO}
\label{sec_ephBY}



\begin{figure}
\begin{center}
\includegraphics[height=0.80\textwidth]{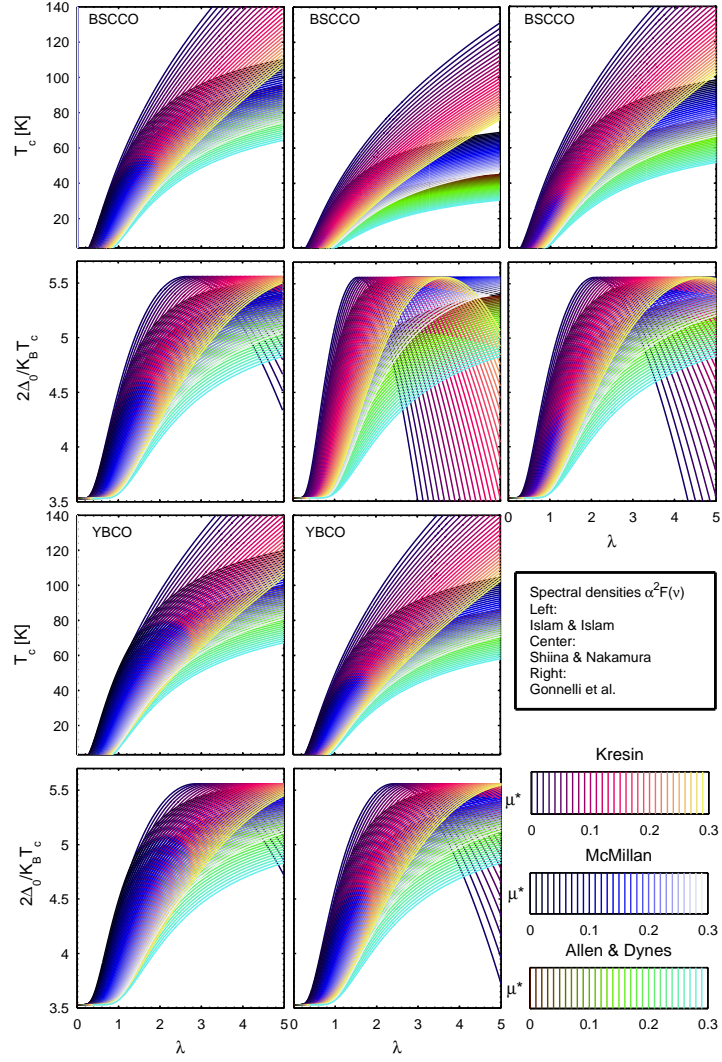} 
\end{center}
\caption{\label{Fig_8}  Plots of the critical temperatures $T_{c}$ (top), the ratio gap $2\Delta_{0}/k_{B}T_{c}$ (middle) and the gap $\Delta_{0}$ (bottom) for BSCCO (top) and YBCO (bottom),  all of them represented as functions of the electron-phonon coupling parameter $\lambda$ and the Coulomb pseudopotential $\mu^{*}$. $\lambda$ has been determined from the spectral densities of Islam \& Islam~\cite{Islam00} (left), Shiina \& Nakamura~\cite{Shiina90} (centre), and Gonnelli et al.~\cite{Gonnelli98} (right) in the specific case of BSCCO (fig.~\ref{Fig_1}). We have used three different approximations, the Kresin's formula~\cite{Kresin87}, the McMillan's formula~\cite{McMillan68}, and the  Allen-Dynes formula~\cite{Allen75}.}
\end{figure}

\begin{figure}
\begin{center}
\includegraphics[height=.8\textwidth]{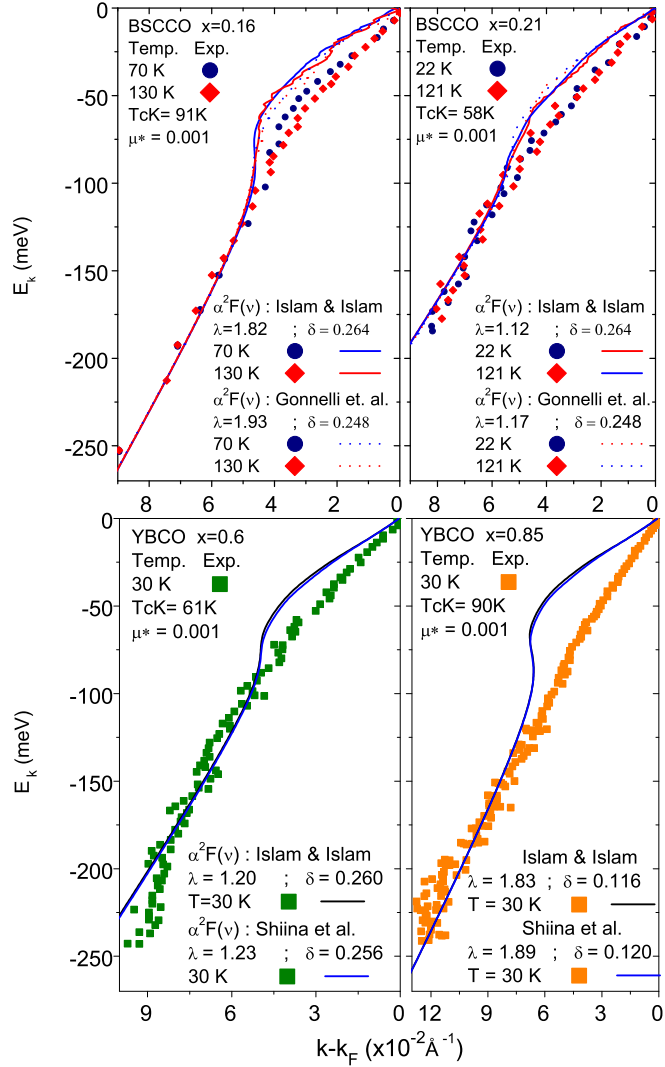} 
\end{center}
\caption{\label{Fig_9}  The renormalised energy $E_{k}$ as a function of the momentum $k-k_{F}$ under the assumption that the nodal dispersion is a consequence of the electron-phonon coupling and the Coulomb effects are completely contained in the bare energies. Then, we assume the very weak Coulomb pseudopotential $\mu^*=0.001$ within the more favourable approximation for the calculus of $T_{c}$ within the phonon hypothesis, i.e. the Kresin Formula (Fig.~\ref{Fig_8}) and the EPI spectral density $\alpha^2\emph{F}(\omega)$ obtained by the methods of Refs.~\cite{Islam00,Gonnelli97} in BSCCO cases (top), and Refs.~\cite{Islam00,Shiina90} in YBCO cases (bottom).}
\end{figure}

The analysis of the results for the BSCCO OPD90 and BSCCO OVD85 cases, already revealed in  previous results (Ref.~\cite{Ruiz09}) that the influence of the electron-phonon mechanism is seemingly weak in spite of the fact that the kink effect is reproduced within this scenario. In fact, independently of the method used for obtaining the electron-phonon spectral density (Fig.~\ref{Fig_1}) the maximal values for the electron-phonon coupling parameter which reproduce the kink structure in BSCCO OPD90 ($\lambda \simeq 1.3$) and OVD58 ($\lambda \simeq 0.85$) samples (Fig.~\ref{Fig_6}), both underestimate the experimental critical temperatures in around of $56\%$ and $67\%$ respectively. This value has been obtained by assuming a conventional value for the Coulomb pseudopotential $(\mu^{*}=0.13)$ (Fig.~\ref{Fig_2}). Now, recalling the impossibility of describing a satisfactory perturbation theory for the Coulomb effects, here we have analysed the influence of $\mu^{*}$ on the $T_{c}$ approximations to determine possible enhancements of the thermodynamic properties as functions of the electron-phonon coupling parameter $\lambda$ and their corresponding spectral densities (Fig.~\ref{Fig_8}). In particular, taking the value $\mu^{*}=0.001$ instead of assuming the conventional $\mu^{*}=0.13$, the Coulomb effects are even more negligible in the renormalised energy. Thus, in BSCCO OPD90 one obtains an enhancement of around $30\%$ ($T_{c}\sim 68 K$) as compared to the conventional $T_{c}$ value ($T_{c}\sim 40 K$). This fact is even more notorious in BSCCO OVD58 where $T_{c}$ improves around $37\%$ ($T_{c}\sim 41 K$). However, even when favoured, the phonon mechanism does not seem to be the only mechanism to be considered in the pair formation that leads at the higher critical temperatures observed in BSCCO OPD90 and OVD58 samples. Notice that, in order to show this, we have assumed the most favourable scenario for the phonon hypothesis, i.e.: {\bf 1}. we have considered the methods of Islam \& islam (Ref.~\cite{Islam00}) and Gonnelli et al. (Ref.~\cite{Gonnelli98}) to determine electron-phonon spectral densities. {\bf 2}. We have assumed a very weak Coulomb pseudopotential ($\mu^{*}=0.001$), and {\bf 3}. The empirical formula of Kresin to determine $T_{c}$ has been used (see figures~\ref{Fig_1}~\&~\ref{Fig_8}). Within this scenario, we get $\lambda_{Islam}(T_{c}^{Kresin}=91K)\simeq1.82$ and $\lambda_{Gonnelli}(T_{c}^{Kresin}=91K)\simeq1.93$, with the renormalisation parameters for the ARPES ``bare'' dispersion $\delta=0.264$ and $\delta=0.248$ respectively  for BSCCO OPD90. Regarding the case of BSCCO OVD58, the electron-phonon coupling parameters obtained are $\lambda_{Islam}(T_{c}^{Kresin}=58K)\simeq1.12$ and $\lambda_{Gonnelli}(T_{c}^{Kresin}=58K)\simeq1.17$, with the same $\delta$ values. Then, we observe a widening of the kink effect suggesting the existence of at least one additional perturbation mechanism reducing the momentum of the dispersed quasiparticles (see Fig.~\ref{Fig_9}). 


There are several possibilities as candidates of additional perturbation mechanisms, a suitable one being to investigate hidden factors in the intensity of the magnetic resonance mode and its evolution with the doping level~\cite{Zhou05,Park08,Graf07,Wilson06,Zhao07,Hwang04}.


Now, we will consider the same hypotheses dealed above for the BSCCO family, i.e.: improving the phonon mechanism by assuming a very weak contribution of the Coulomb pseudopotential ($\mu^{*}=0.001$), in an effort to obtain appropriate values of $T_{c}$ in the case YBCO UD35, with the same value of the electron-phonon coupling $\lambda=0.80$. Going into detail, we have used the methods by Islam \& Islam~\cite{Islam00} and Shiina \& Nakamura~\cite{Shiina90} to determine the electron-phonon spectral densities (Fig.~\ref{Fig_1}), and then different approaches for $T_{c}$ have been evaluated (Fig.~\ref{Fig_2}). Thus, with the method from Ref.\cite{Islam00} we obtain: $T_{c}^{Kresin}\simeq37.69$, $T_{c}^{McMillan}\simeq40.04$, and $T_{c}^{Allen-Dynes}\simeq33.25$, and from the more rigorous method of Ref.\cite{Shiina90} we obtain: $T_{c}^{Kresin}\simeq36.86$, $T_{c}^{McMillan}\simeq35.01$, and $T_{c}^{Allen-Dynes}\simeq28.71$. Then, under the assumption that the Coulomb effects don't play any important role on the electron self-energy, one can conclude that the phonon mechanism could be the most relevant interaction mechanism in the YBCO UD35 sample. However, regarding the $T_{c}$ values of the under-doped family, they are still underestimated by a percentage of $62\%$ in YBCO UD61 and around $80\%$ in YBCO OVD90.


On the other side, in Fig.~\ref{Fig_9} we show the maximal influence of the electron-phonon coupling parameters that reproduce $T_{c}$ on the quasiparticle dispersion energies measured by ARPES. Analogously to the BSCCO cases, here we have assumed the most favourable scenario for the phonon hypothesis. We obtain $\lambda_{Islam}(T_{c}^{Kresin}=61K)\simeq1.20$, and $\lambda_{Shiina}(T_{c}^{Kresin}=61K)\simeq1.23$, with the ARPES ``bare'' dispersion renormalised at $\delta=0.260$ and $\delta=0.256$ respectively (Fig.~\ref{Fig_9}). For the over-doped YBCO OVD90, the electron-phonon coupling parameters obtained are $\lambda_{Islam}(T_{c}^{Kresin}=90K)\simeq1.83$ and $\lambda_{Shiina}(T_{c}^{Kresin}=90K)\simeq1.89$, with the ARPES ``bare'' dispersion renormalised at $\delta=0.116$ and $\delta=0.120$ respectively (Fig.~\ref{Fig_9}). Then, the more remarkable widening of the kink effect is again suggesting the existence of at less one additional perturbation mechanism. In fact, in YBCO the importance of an additional mechanism seems to be more significant that in BSCCO, and it could be related to the appearance of a second kink in the under-doped phase. 


Very recent efforts have been focused on the high-energy part of the ARPES data (0.2-1.5 eV) revealing additional changes in the electron energy dispersion ``\textit{waterfalls}'' at $\sim0.4$ eV \cite{Zhang08,Graf08,Chang08,Graf07,Valla07}. The presence of a kink at these high energies immediately raises the question: What type of excitation spectrum is required to produce such renormalisation effect? One possibility is that correlation effects at high orders strongly renormalises the electron-phonon coupling. Moreover, with regard at the photoemission process and the inversion of the data measured in YBCO samples one should take special care with the residual 3-dimensionality and the possible effect of the CuO chains. Nevertheless, these high energy features remain under intense debate \cite{Zhang08,Graf08,Chang08,Graf07,Valla07,Zhu08,Tan08} because it is still unclear whether they represent intrinsic band structure or not \cite{Zhang08}.


\section{CONCLUDING REMARKS}\label{Sec:Conclusions}


In summary, it has been shown that, regarding the ARPES measurements in superconducting cuprates the strong renormalisation of the band structure, customarily related to the dressing of the electron with excitations can be explained in terms of the conventional electron-phonon interaction. In fact, our results suggest that the electron-phonon interaction strongly influences the electron dynamics of the high-$T_{c}$ superconductors, and it is an important mechanism linked with the Fermi surface topology. Thus, the electron-phonon interaction (strong or weak) must be included in any realistic microscopic theory of superconductivity.


Remarkably, the proposal of a simple linear dispersion relation for the electron bare band energy, i.e., $\varepsilon_{k}=(k-k_{F})v_{F<}(1+\lambda^{*})$, with the electronic mass-enhancement parameter $\lambda^{*}$ defined self-consistently by the relation $1+\lambda^{*}=(1-\delta\lambda)^{-1}$ allows to reproduce the appearance of the ubiquitous nodal kink for a wide set of ARPES experiments in the cuprates. Here,  $\lambda$ has been interpreted as the strength of the electron-phonon interaction or electron-phonon coupling parameter, and the dimensionless parameter $\delta$ recognised as a universal property for each family of cuprates. In addition, at the lowest order of the Feynman graphs for the electron-phonon interaction this method reassembles the ``$\lambda$'' values obtained from several discrepant models. 


In this work, we have investigated the influence of the electron-phonon coupling mechanism through different doping levels in several families of HTSC-cuprates. On the one hand, we have evaluated different methods for obtaining the electron-phonon spectral densities and their influence on the electron bare band energy, and on the other hand, several approaches have been recalled to obtain the critical temperatures. Then, our results suggest that at least in the LSCO family, and in the so-called BSCCO UD70 and YBCO UD35 superconductors, the electron-phonon interaction could be the most relevant mechanism involved in the pair formation that leads to the superconductivity. Our conclusion is supported by the experimental evidence of a mass renormalisation of the electronic dispersion curves measured along the nodal direction in ARPES and the reported $T_{c}$ values in good agreement with our theoretical predictions. An excellent comparison between the theory and the available collection of experiments is achieved. Furthermore, we have evaluated the consequences of assuming an enhanced phonon mechanism, through the reduction of the Coulomb's pseudopotential weight. When appropriate $T_c$ values  are obtained by this method, a remarkable widening of the predicted kink effect arises. This fact, suggest that independently of the approximations invoked and even avoiding the influence of the \textit{d}-wave superconducting gap through the nodal ARPES measurements, it doesn't seem possible to elude the existence of additional mechanisms that reduce the momentum of the dispersed quasiparticles in comparison with the phonon mechanism. In this sense, despite the fact that in LSCO the influence of the magnetic mode seems not relevant, it is not possible to ignore its importance over the electron properties of other HTSC families having in mind the recent observation by G. Yu \textit{et al.} \cite{Yu09} on a possible connection between the magnetic excitations and the superconductor gap for a wide range of materials.


\section*{Acknowledgement}

This work was supported by Spanish CICyT project MAT2008-05983-C03-01.
H. S. Ruiz acknowledges a grant from Spanish CSIC (JAE program).
%



\section*{References}

\begin{thebibliography}{60}

\bibitem{Bednorz86}     J. G. Bednorz and K. A. M\"{u}ller, Z. Phys. B \textbf{64}, 18 (1986).
\bibitem{Hufner07}      S. H\"{u}fner (Ed.), \emph{Very High Resolution Photoelectron Spectroscopy} (Springer, Berlin, 2007).
\bibitem{Kulic00}	M.L. K\'{u}lic, Phys. Rep. 338, 1 (2000).
\bibitem{Smith01}	N. V. Smith, Phys. Rev. B \textbf{64}, 155106 (2001).
\bibitem{Kulic04}      M. L. Kuli\'{c}, \emph{Lectures on the physics of highly correlated electron systems VIII: 8th Training Course in the Physics of Correlated Electron Systems and High-Tc Superconductors. AIP Conference Proceedings}, Volume 715, pp. 75-158 (2004).
\bibitem{Devereaux04}   T. P. Devereaux, T. Cuk, Z.-X. Shen, and N. Nagaosa, Phys. Rev. Lett. \textbf{93}, 117004 (2004).
\bibitem{Lanzara01}     A. Lanzara, P. V. Bogdanov, X. J. Zhou, S. A. Kellar, D. L. Feng, E. D. Lu, T. Yoshida, H. Eisaki, A. Fujimori, K. Kishio, J.-I. Shimoyama, T. Noda, S. Uchida, Z. Hussain, and Z.-X. Shen, Nature \textbf{412}, 510 (2001).
\bibitem{Zhou03}        X. J. Zhou, T. Yoshida, A. Lanzara, P. V. Bogdanov, S. A. Kellar, K. M. Shen, W. L. Wang, F. Ronning, T. Sasagawa, T. Kakeshita, T. Noda, H. Eisaki, S. Uchida, C. T. Lin, F. Zhou, J. W. Xiong, W. X. Ti, Z. X. Zhao, A. Fujimori, Z. Hussain, and Z.-X. Shen, Nature \textbf{423}, 398 (2003).
\bibitem{Zhou07}        X. J. Zhou, T. Cuk, T. Devereaux, N. Nagaosa, and Z.-X. Shen \emph{Handbook of High-Temperature Superconductivity} (Springer, New York, 2007), chap. 3, p. 87.
\bibitem{Zhou02}        X. J. Zhou, Z. Hussain, and Z.-X. Shen, J. Electron Spectrosc. Relat. Phenom. \textbf{126}, 145 (2002).
\bibitem{Zhou05}        X. J. Zhou, J. Shi, T. Yoshida, T. Cuk, W. L. Yang, V. Brouet, J. Nakamura, N. Mannella, S. Komiya, Y. Ando, F. Zhou, W. X. Ti, J. W. Xiong, Z. X. Zhao, T. Sasagawa, T. Kakeshita, H. Eisaki, S. Uchida, A. Fujimori, Z. Zhang, E. W. Plummer, R. B. Laughlin, Z. Hussain, and Z.-X. Shen, Phys. Rev. Lett. \textbf{95}, 117001 (2005).
\bibitem{Kordyuk06}     A. A. Kordyuk, S. V. Borisenko, V. B. Zabolotnyy, J. Geck, M. Knupfer, J. Fink, B. B\"{u}chner, C. T. Lin, B. Keimer, H. Berger, A. V. Pan, S. Komiya, and Y. Ando, Phys. Rev. Lett. \textbf{97}, 017002 (2006).
\bibitem{Xiao07}        Y. X. Xiao, T. Sato, K. Terashima, H. Matsui, T. Takahashi, M. Kofu, and K. Hirota, Physica C \textbf{463-465}, 44 (2007).
\bibitem{Takahashi07}   T. Takahashi, Physica C \textbf{460-462}, 198 (2007).
\bibitem{Gweon04}       G.-H. Gweon, T. Sasagawa, S. Y. Zhou, J. Graf, H. Takagi, D.-H. Lee, and A. Lanzara, Nature \textbf{430}, 187 (2004).
\bibitem{Douglas07}     J. F. Douglas, H. Iwasawa, Z. Sun, A. V. Fedorov, M. Ishikado, T. Saitoh, H. Eisaki, H. Bando, T. Iwase, A. Ino, M. Arita, K. Shimada, H. Namatame, M. Taniguchi, T. Masui, S. Tajima, K. Fujita, S. Uchida, Y. Aiura, and D S. Dessau, Nature \textbf{446}, E5 (2007).
\bibitem{Johnson01}     P. D. Johnson, T. Valla, A. V. Fedorov, Z. Yusof, B. O. Wells, Q. Li, A. R. Moodenbaugh, G. D. Gu, N. Koshizuka, C. Kendziora, Sha Jian, and D. G. Hinks, Phys. Rev. Lett \textbf{87}, 177007 (2001).
\bibitem{Borisenko06}   S. V. Borisenko, A. A. Kordyuk, V. Zabolotnyy, J. Geck, D. Inosov, A. Koitzsch, J. Fink, M. Knupfer, B. B\"{u}chner, V. Hinkov, C. T. Lin, B. Keimer, T. Wolf, S. G. Chiuzb\u{a}ian, L. Patthey, and R. Follath, Phys. Rev. Lett \textbf{96}, 117004 (2006).
\bibitem{Zhang08} 	W. Zhang, G. Liu, J. Meng, L. Zhao, H. Liu, X. Dong, W. Lu, J. S. Wen, Z. J. Xu, G. D. Gu, T. Sasagawa, G. Wang, Y. Zhu, H. Zhang, Y. Zhou, X. Wang, Z. Zhao, C. Chen, Z. Xu, and X. J. Zhou, Phys. Rev. Lett. \textbf{101}, 017002 (2008).
\bibitem{Graf08}   J. Graf, M. d'Astuto, C. Jozwiak, D. R. Garcia, N. L. Saini, M. Krisch, K. Ikeuchi, A. Q. R. Baron, H. Eisaki, and A. Lanzara, Phys. Rev. Lett \textbf{100}, 227002 (2008).
\bibitem{Reznik08}      D. Reznik, G. Sangiovanni, O. Gunnarsson, and T. P. Devereaux, Nature \textbf{455}, E6 (2008).
\bibitem{Giustino08}    F. Giustino, M. L. Cohen, and S. G. Louie, Nature \textbf{452}, 975 (2008).
\bibitem{Park08}        S. R. Park, D. J. Song, C. S. Leem, C. Kim, C. Kim, B. J. Kim, and H. Eisaki, Phys. Rev. Lett. \textbf{101}, 117006 (2008).
\bibitem{Chang08}      J. Chang, M. Shi, S. Pailh\'{e}s, M. M\aa{a}nsson, T. Claesson, O. Tjernberg, A. Bendounan, Y. Sassa, L. Patthey, N. Momono, M. Oda, M. Ido, S. Guerrero, C. Mudry, and J. Mesot, Phys. Rev. B \textbf{78}, 205103 (2008).
\bibitem{Dahm09}    T. Dahm, V. Hinkov, S. V. Borisenko, A. A. Kordyuk, V. B. Zabolotnyy, J. Fink, B. B\"{u}chner, D. J. Scalapino, W. Hanke, B. Keimer, Nature Physics 5, 217 (2009).
\bibitem{Graf07}   J. Graf, G.-H. Gweon, K. McElroy, S. Y. Zhou, C. Jozwiak, E. Rotenberg, A. Bill, T. Sasagawa, H. Eisaki, S. Uchida, H. Takagi, D.-H. Lee, and A. Lanzara, Phys. Rev. Lett \textbf{98}, 067004 (2007).
\bibitem{Terashima06}   K. Terashima, H. matsui, D. Hashimoto, T. Sato, T. Takahashi, H. Ding, T. Yamamoto and K. Kadowaki, Nature Physics \textbf{2}, 27 (2006).
\bibitem{Wilson06}    S. D. Wilson, P. Dai, S. Li, S. Chi, H. J. Kang, and J. W. Lynn, Nature \texttt{442}, 59 (2006).
\bibitem{Zhao07}       J. Zhao, P. Dai, S. Li, P. G. Freeman, Y. Onose, and Y. Tokura, Phys. Rev. Lett. \textbf{99}, 017001 (2007).
\bibitem{Sato03}       T. Sato, H. Matsui, T. Takahashi, H. Ding, H.-B. Yang, S.-C. Wang, T. Fujii, T. Watanabe, A. Matsuda, T. Terashima, and K. Kadowaki, Phys. Rev. Lett. \textbf{91}, 157003 (2003).
\bibitem{Bardeen57}	J. Bardeen, L. N. Cooper y J. R. Schrieffer, Phys. Rev. \textbf{106} 162-, \textbf{108} 1175-(1957).
\bibitem{Schrieffer07}\emph{Handbook of High-Temperature Superconductivity: Theory and Experiment}, edited by J. R. Schrieffer and J. S. Brooks (Springer, New York, 2007).
\bibitem{Carbotte90}	J. P. Carbotte, Rev. Mod. Phys. \textbf{62}, 1027 (1990).
\bibitem{Ruiz09}        H. S. Ruiz and A Bad\'{\i}a Maj\'{o}s, Phys. Rev. B \textbf{79}, 054528 (2009).
\bibitem{Allen78}       P. B. Allen, Phys. Rev. B \textbf{18}, 5217 (1978).
\bibitem{Allen76}	    P. B. Allen, Phys. Rev. B 13, 1416 (1976).
\bibitem{Allen82}       P. B. Allen and B. Mitrovi\'c, in \emph{Solid State Physics}, edited by H. Ehrenreich, F. Seitz, and D. Turnbull (Academic, New York, 1982), Vol. 37, p. 1.
\bibitem{Ashcroft76}    N. W. Ashcroft and N. D. Mermin, \emph{Solid State Physics} (Saunders College Publishing, 1976), chap. 26, p. 521.
\bibitem{Doniach98}	S. Doniach, and E. H. Sondheimer. Green's Functions for Solid State Physicists, (Imperial College Press. Singapore, 1998).
\bibitem{Grimvall81}	G. Grimvall, Sel. Top. Solid State Phys. \textbf{16} (1981).
\bibitem{Gonnelli98}    R. S. Gonnelli, G. A. Ummarino, and V. A. Stepanov, J. Phys. Chem. Solids \textbf{59}, 2058 (1998).
\bibitem{Islam00}       A. T. M. N. Islam and A. K. M. A. Islam, J. Supercond. \textbf{13}, 559 (2000).
\bibitem{Shiina90}      Y. Shiina and Y. O. Nakamura, Sol. State Commun. \textbf{76}, 1189 (1990).
\bibitem{Renker87}      B. Renker, F. Gompf, E. Gering, N. N\"{u}cker, D. Ewert, W. Reichardt, and H. Rietschel, Z. Phys. B \textbf{67}, 15 (1987).
\bibitem{Renker88}      B. Renker, F. Gompf, E. Gering, D. Ewert, and A. Dianoux, Z. Phys. B \textbf{73}, 309 (1988).
\bibitem{Renker89}      B. Renker, F. Gompf, D. Ewert, P. Adelmann, H. Schmidt, E. Gering, and H. Mutka, Z. Phys. B \textbf{77}, 65 (1989).
\bibitem{Gonnelli97}    R. S. Gonnelli, G. A. Ummarino, and V. A. Stepanov, Physica C \textbf{275}, 162 (1997).
\bibitem{Ummarino97}    G. A. Ummarino, R. S. Gonnelli, and V. A. Stepanov, Physica C \textbf{282}, 1501 (1997).
\bibitem{McMillan68}    W. L. McMillan, Phys. Rev. \textbf{167}, 331 (1968).
\bibitem{Allen75}       P. B. Allen and R. C. Dynes, Phys. Rev. B \textbf{12}, 905 (1975).
\bibitem{Kresin87}	V. Kresin, Phys. Lett. A \textbf{122}, 434 (1987); S. A. Wolf and V. Z. Kresin, in \textit{Proc. of the International Workshop on Anharmonic Properties of High-$T_{c}$ cuprates}, D. Mihailovic et al., ed. (World Scientific, Singapour, 1995) p. 232.
\bibitem{Berk66}     N. F. Berk and J. R. Schrieffer, Phys. Rev. Lett. \textbf{17}, 433 (1966).
\bibitem{Hwang04}       J. Hwang, T. Timusk, and G. D. Gu, Nature \textbf{427}, 714 (2004).
\bibitem{Damascelli03}	A. Damascelli, Z. Hussain, and Z.-X. Shen
Rev. Mod. Phys. \textbf{75}, 473 (2003)
\bibitem{Uchida87}      S. Uchida, H. Takagi, K. Kishio, K. Kitazawa, K. Fueki, and S. Tanaka, Jpn. J. Appl. Phys. \textbf{26}, L443 (1987).
\bibitem{Dietrich87}    M. R. Dietrich, W. H. Fietz, J. Ecke, B. Obst and C. Politis, Z. Phys. B \textbf{66}, 283 (1987).
\bibitem{Heid08}        R. Heid, K.-P Bohnen, R. Zeyher, and D. Manske Phys. Rev. Lett. \textbf{100}, 137001 (2008).
\bibitem{Mattheiss87}   L. F. Mattheiss, Phys. Rev. Lett. \textbf{58}, 1028 (1987).
\bibitem{Weber87}       W. Weber, Phys. Rev. Lett. \textbf{58}, 1371 (1987).
\bibitem{Weber88}       W. Weber, Phys. Rev. B. (Rapid Comm.) \textbf{37}, 599 (1988).
\bibitem{Valla07}      T. Valla, T. E. Kidd, W.-G. Yin, G. D. Gu, P. D. Johnson, Z.-H. Pan, and A. V. Fedorov, Phys. Rev. Lett. \textbf{98}, 167003 (2007).
\bibitem{Zhu08}     Lijun Zhu, Vivek Aji, Arkady Shekhter, and C. M. Varma, Phys. Rev. Lett. \textbf{100}, 057001 (2008).
\bibitem{Tan08}      F. Tan and Q.-H. Wang, Phys. Rev. Lett. \textbf{100}, 117004 (2008).
\bibitem{Yu09}       G. Yu, Y. Li, E. M. Motoyama and M. Greven, Nature Physics \textbf{5}, 873 (2009).

\end{thebibliography}
\end{document}